\documentclass[10pt,aps,prx,twocolumn, nofootinbib,superscriptaddress]{revtex4-1}

\usepackage{natbib}
\usepackage{aas_macros} 
\usepackage{amssymb,amsmath,accents}
\usepackage{caption} 
\usepackage{subcaption}
\usepackage{verbatim}
\usepackage{ragged2e}
\let\oldjustifying\justifying
\DeclareRobustCommand{\justifying}{\oldjustifying}
\usepackage{graphicx}
\usepackage{color,units}
\usepackage[dvipsnames]{xcolor}
\usepackage{soul}
\usepackage{multirow}
\usepackage{array, makecell, tabularx, cellspace}
\newcolumntype{L}{>{\raggedright\arraybackslash}X}
\newcolumntype{C}{>{\centering\arraybackslash}X}
\usepackage{booktabs} 
\usepackage{bm}
\usepackage{hyperref}
\usepackage{listings}
\usepackage[normalem]{ulem}
\hypersetup{
    unicode=false,          
    pdftoolbar=true,        
    pdfmenubar=true,        
    pdffitwindow=false,     
    pdfstartview={FitH},    
    pdfauthor={Boris Goncharov},     
    colorlinks=true,       
    linkcolor=BrickRed,          
    citecolor=BrickRed,        
    urlcolor=RoyalBlue
}
\graphicspath{{figures/}}

\begin{document}

\definecolor{dkgreen}{rgb}{0,0.6,0}
\definecolor{gray}{rgb}{0.5,0.5,0.5}
\definecolor{mauve}{rgb}{0.58,0,0.82}

\lstset{frame=tb,
  	language=Matlab,
  	aboveskip=3mm,
  	belowskip=3mm,
  	showstringspaces=false,
  	columns=flexible,
  	basicstyle={\small\ttfamily},
  	numbers=none,
  	numberstyle=\tiny\color{gray},
 	keywordstyle=\color{blue},
	commentstyle=\color{dkgreen},
  	stringstyle=\color{mauve},
  	breaklines=true,
  	breakatwhitespace=true
  	tabsize=3
}
\title{A Joint Optimal Search for Gravitational Waves from Resolved and Unresolved Supermassive Binary Black Holes with Pulsar Timing Arrays}
\author{Boris Goncharov}%
 \email{boris.goncharov@aei.mpg.de}
\affiliation{Max Planck Institute for Gravitational Physics (Albert Einstein Institute), 30167 Hannover, Germany}
\affiliation{Leibniz Universität Hannover, 30167 Hannover, Germany}
\author{Gabriela Sato-Polito}
\affiliation{School of Natural Sciences, Institute for Advanced Study, Princeton, NJ 08540, United States}
\author{Xiaoming Bi}
\affiliation{Max Planck Institute for Gravitational Physics (Albert Einstein Institute), 30167 Hannover, Germany}
\affiliation{Leibniz Universität Hannover, 30167 Hannover, Germany}
\author{Matias Zaldarriaga}
\affiliation{School of Natural Sciences, Institute for Advanced Study, Princeton, NJ 08540, United States}

\date{\today}

\begin{abstract}

We introduce, from first principles, a joint model of the gravitational wave background (GWB) and brightest supermassive black hole binary (SMBHB) sources that may be individually resolvable in Pulsar Timing Array (PTA) searches for gravitational waves. 
We propose the characteristic number of SMBHB sources, $N_{\rm c}$, as a detection statistic for the astrophysical origin of the GWB. 
We then demonstrate how the brightest SMBHBs assist in resolving $N_{\rm c}$. 
Applying our method to the simulated NANOGrav 15-year data, which replicates all aspects of real data's known noise, observations, and the inferred GWB power spectrum, we demonstrate direct astrophysical limits on the strain amplitude of individually resolvable SMBHBs. 
We find that 21 of 114 SMBHB candidates from active galactic nuclei observations are in tension with NANOGrav's observations. 
In contrast, only one candidate is in tension with the NANOGrav data based on the upper limits reported in the original analysis. 
Constraining the Poisson-specific characteristic number of SMBHBs, $N_{\rm c}$, at ${\rm yr}^{-1}$, we outline implications for the population properties of SMBHBs. 
Based on our new model applied to the simulated NANOGrav data, we calculate the probability of detecting GWs from isolated SMBHB in the 15-year data to be 2\% at the ${\rm SNR}=5$ level. 
Our projection towards the expected NANOGrav 20-year data suggests an increase to 5\%. 
With this, we estimate the probability of finding an outlier with an SNR of 2 in the upcoming NANOGrav 20-year data to be $40\%$.

\end{abstract}

\maketitle

\section{Introduction}

Pulsar Timing Array (PTA) experiments have reported evidence for the gravitational wave background (GWB) at nanohertz frequencies~\cite{NG15_GWB,EPTA_DR2_GW,PPTA_DR3_GWB,CPTA_DR1_GWB,MT_DR1_GW,NG12_GWB, GoncharovShannon2021, ChenCaballero2021, IPTA_DR2_GWB,GoncharovThrane2022}. 
The most expected, astrophysical, GWB emerges from the stochastic incoherent superposition of strain from adiabatically inspiralling supermassive black hole binaries (SMBHBs)~\cite{JaffeBacker2003}. 
Various studies have shown that such an origin is expected to reveal inconsistencies with assumptions employed for the initial GWB detection: the isotropic angular power spectrum, the Gaussian process description, and the power-law strain spectrum~\cite{SesanaVecchio2010,MingarelliSidery2013,RosadoSesana2015}. 
The inconsistencies arise due to the discrete point-source nature of the binaries that constitute the GWB, and they are the reason for the predicted detection of individual SMBHBs as a continuous gravitational wave (CW) to follow. 
When solidified in a self-consistent, testable model, deviations from the isotropic, Gaussian GWB model could enable a smoking gun identification of the origin of the GWB -- an important milestone that follows the ongoing detection and confirmation of the signal. 

A contemporary data analysis treatment of one astrophysical stochastic process seeding both the background of gravitational waves (GWs) and individually resolvable CW signals remains fragmented. 
In particular, collaboration-led searches for the background~\cite{NG15_GWB} and the resolvable sources~\cite{NG15_SMBHB} are performed as separate studies. 
Whereas the priors on the strain amplitude of CWs and the characteristic strain spectrum of the GWB in these analyses are chosen to be uninformative. 
The inference of population properties of SMBHBs based on PTA data is typically a separate analysis, usually linking SMBHB population models to the overall GWB amplitude or spectral shape rather than jointly modeling the unresolved background and individually resolvable sources in one likelihood~\cite{KelleyBlecha2017,MiddletonChen2018,ChenSesana2019,NG15_HOLODECK}.
This remains a robust strategy for the initial detection and characterization of these signals, but potentially suboptimal and insufficient for a complete science extraction from future PTA observations. 
Searches for deviations from the isotropic angular power distribution of the GWB and aims to resolve features in the strain spectrum of specific SMBHB models remain useful probes of the multifaceted nature of the astrophysical GWB, but they still lack a definitive angle. 
Just like the Hellings-Downs correlations~\cite{HellingsDowns1983} are a definitive signature of the tensor metric perturbations of GWs of General Relativity, a mathematical model of GWs from SMBHBs is a key to unlocking the detection and characterization of GWBs' astrophysical nature. 

Perhaps the most fundamental feature of the GW signal from SMBHBs is the transition from the Gaussian large-number statistics to the Poissonian small-number statistics. 
Sato-Polito and Zaldarriaga~\cite{Sato-PolitoZaldarriaga2025} developed a model of this transition for SMBHBs in circular orbits that lose their orbital energy solely by the emission of GWs. 
They showed that two SMBHB population parameters --- the mass scale and the number density of SMBHBs, which can also be recast as the characteristic number of sources and the characteristic strain amplitude -- determine the astrophysical probability density function (PDF) of the total characteristic GW strain from such SMBHBs. 
This PDF is an approximation, but PTAs are unlikely to distinguish it from a more comprehensive model in the foreseeable future. 
The characteristic number of sources, termed $N_{\rm c}$, is the number of SMBHBs required to produce the background's characteristic strain given a population model. 
The number of SMBHBs contributing to every GW frequency is modelled as a Poissonian process. 
As the characteristic number of sources increases, which scales with frequency, the PDF of the total characteristic strain becomes thinner, approaching the standard astrophysical model for SMBHB: a Gaussian process with the characteristic strain spectrum given by a power law with the spectral index of $-2/3$. 
When the characteristic number of sources approaches one source or less, the finite-width PDF of the total characteristic strain leads to visible stochastic fluctuations in the strain spectrum with respect to the characteristic strain corresponding to the mean of the density function. 
The same statistical property of GWs from SMBHBs has been explored from another angle by Xue, Pan, and Dai~\cite{XuePan2025}, additionally considering the effect of the interference of GWs from different SMBHBs. 
The effect of source interference on Hellings-Downs correlations is explored by Wu, Bi, and Huang~\cite{WuBi2024}. 
Earlier, the calculation has also been performed numerically by Ellis~\textit{et al}~\cite{EllisFairbairn2023}. 
Lamb and Taylor~\cite{LambTaylor2024} calculated the scaling of the moments of the total characteristic strain PDF with frequency. 
Hisamatsu and Kyutoku extend this approach, finding the consistency relation between the kurtosis and the skewness of the PDF, and expanding on the astrophysics extraction by studying the mass and redshift dependency of the model~\cite{HisamatsuKyutoku2026}. 
Falxa and Sesana~\cite{FalxaSesana2026} proposed the approach of fitting astrophysical signals to Gaussian mixture models that allow the detection of SMBHBs' non-Gaussian signatures. 
Recently, Ali-Ha\"imoud~\cite{Ali-Haimoud2026} derived a universal PDF for the characteristic strain distribution in the limit of large $N_c$, pointing out that the log-Normal approximation for the total characteristic strain used in Ref.~\cite{NG15_HOLODECK} is not correct. 
One of the earliest efforts by Sardesai, Simon, and Vigeland~\cite{SardesaiSimon2024} introduces a $t$-process description for deviations from the power-law strain spectrum of the GWB, but it is also a purely phenomenological approximation. 

In this work, building on Ref.~\cite{Sato-PolitoZaldarriaga2025}, we introduce a definitive testable model of a complete astrophysical GW signal including both fully resolved and completely unresolved sources, as well as the end-to-end study of this model with the simulated NANOGrav 15-year data~\cite{NG15_data}.
In particular, we extend the calculation of Ref.~\cite{Sato-PolitoZaldarriaga2025} by introducing the PDF of the total characteristic strain of the brightest SMBHB source. 
Incorporation of the brightest source PDF for CW searches for individual SMBHBs enables a marginalization over GW interference and isotropy deviations for weak signals, as well as over potential limitations of inclination averaging pointed out in~\cite{Ali-Haimoud2026}. 
Because the astrophysical model is more informative and rooted in physics, it may potentially allow more stringent observational and astrophysical constraints. 
Furthermore, whereas the analysis of PTA data in Ref.~\cite{Sato-PolitoZaldarriaga2025} is based on refitting of the GWB power spectrum inferred from the NANOGrav data, we develop a complete data analysis methodology around their total characteristic strain calculation in a global fit of the PTA data to all signals and noise. 
We test the new model and the new approach for the case of SMBHBs in circular orbits evolving in frequency beyond the PTA observational timescale and entirely due to GW emission, but it remains ready to be extended to eccentric binaries and environmental effects. 

The rest of the paper is organized as follows. 
In Section~\ref{sec:model}, we introduce our joint hierarchical model of the total characteristic strain probability density for the GW background and the brightest source. 
In Section~\ref{sec:methods}, we summarize the data analysis methodology. 
In Section~\ref{sec:phenom}, we outline the observations and the astrophysical capabilities of our model on the simulated PTA data.
In Section~\ref{sec:results}, we present the results of our analysis of the simulated NANOGrav 15-year data.
In Section~\ref{sec:astro}, we discuss the astrophysical properties of SMBHBs based on our results.
In Section~\ref{sec:conclusion}, we conclude the findings.

\section{\label{sec:model}The Astrophysical Model of Nanohertz Gravitational Waves}

In this Section, we outline contributions to the characteristic GW strain from SMBHBs at every frequency: the total strain $h_{\rm t}$, the strain of a single SMBHB source $h_{\rm s}$, the strain of the brightest source $h_{\rm cw}$, and the strain of all sources except the brightest $h_{\rm t-1}$. 
Characteristic strain is defined as $h_{\rm c}(f) \equiv \sqrt{S(f)f}$, where $S(f)$ is the strain power spectral density and $f$ is the GW frequency. 
In practice, because the brightest sources are the first ones that may become individually resolvable, they are treated as deterministic directional signals (continuous waves, hence denoted ``cw''), whereas the remaining sources ($h_{\rm t-1}$) are treated as an isotropic, stochastic signal. 
For the cases where none of the SMBHBs are individually resolvable, all GW signals can be treated as the latter ($h_{\rm t}$). 

We start by examining the expression for the PDF of $h_{\rm t}$, expanding Equation 18 of Ref.~\cite{Sato-PolitoZaldarriaga2025} representing an inverse Fourier transform:
\begin{equation}
p(h_{\rm t}^2) = \int\ \exp\left\{i \omega h_{\rm t}^2 \left(\mathcal{F}\left[\frac{dN}{dh_{\rm s}^2}\right] - \bar N\right)\right\} d\omega.
\label{eq:p_h_tot}
\end{equation}
Here, $\mathcal{F}$ is the Fourier transform applied to the GW luminosity function $dN/dh_{\rm s}^2$, $\omega$ is the Fourier variable for $h_{\rm s}^2$, and $\bar N$ is the full integral of the luminosity function over $h_{\rm s}^2$. 
We define the characteristic strain \textit{amplitude}, $h_{\rm c}$, a widely used quantity in the PTA literature, as the mean of $p(h_{\rm t}^2)$.
This way, it follows the power law $h_{\rm c}(f)=A(f~{\rm yr})^{-2/3}$ in GW frequency for SMBHBs evolving adiabatically due to emission of GWs in circular orbits. 
In the limit of the infinite number of SMBHBs, $p(h_{\rm t}^2) = \delta(h_{\rm t}^2)$, and $h_{\rm t}(f) = h_{\rm c}(f)$. 

The average characteristic strain is given by
\begin{equation}
    h_c^2(f) = \int d\log h^2_s \frac{dN}{d\log h^2_s} h^2_s.
\end{equation}
The kernel in the equation above is peaked at a value $h^2_{\rm s,peak}$, and we define the rescaled variable $x=h^2_{\rm s}/h^2_{\rm s,peak}$. Ref.~\cite{Sato-PolitoZaldarriaga2025} identified that quantity
\begin{equation}
\mu(x) = \frac{1}{N_{\rm c}}\frac{dN}{d\log x}, 
\label{eq:mu_x}
\end{equation}
is (quasi-)invariant under any SMBHB population model, which we discuss in detail in App.~\ref{app:model}. Here, $N_{\rm c} \equiv h^2_{\rm c}/h^2_{\rm s,peak}$ is the characteristic number of sources, which scales with frequency as $N_{\rm c}(f) \propto f^{-11/3}$, since $h^2_{\rm s,peak}(f) \propto f^{7/3}$ and $h^2_c \propto f^{-4/3}$. We use this invariance to evaluate Equation~\ref{eq:p_h_tot} numerically. 
In this work, we calculate $\mu(x)$ as prescribed by Ref.~\cite{Sato-PolitoZaldarriaga2025} and linearly interpolate.
Additionally, it is possible to treat $\mu(x)$ as the log-Normal approximation to a high precision. 
The procedure for the latter is also described in Appendix~\ref{app:model:mu_x_logNormal}.

From the invariance of $\mu$ with respect to SMBHB population properties, it follows that $p(h_{\rm t}^2)$ is a function of only two variables, $N_{\rm c}$ and $h_{\rm c}$, the scaling of which across frequency is known. 
Thus, we calculate $(N_{\rm c},h_{\rm c})$ only for the referenced GW frequency $f_{\rm ref}$ and rescale the PDF to other frequencies. 
Throughout this work, we chose $f_{\rm ref}={\rm yr}^{-1}$. 
In terms of Bayesian inference of $(h_{\rm t},h_{\rm cw},h_{\rm c})$, $p(h^2)$ is referred to as the prior distribution --- a model of parameters' distribution provided by nature. 
With this, $(N_{\rm c},h_{\rm c})$ are referred to as astrophysical hyperparameters. 
So, when performing data analysis, we denote $p(h_{\rm t}^2,h_{\rm cw}^2,h_{\rm t-1}^2)=\pi(h_{\rm t}^2,h_{\rm cw}^2,h_{\rm t-1}^2|N_{\rm c},h_{\rm c})$, where $\pi$ refers to the prior and the condition refers to a parametrization of the prior of hierarchical Bayesian inference. 

In this work, we introduce PDFs for $h_{\rm cw}$ and the remainder $h_{\rm t-1}$. 
The PDF of $h_{\rm cw}$, which is a maximum value of $h_{\rm s}$, is given by 
\begin{equation}
p(h_{\rm s}) = h_{\rm s,peak}^2 \frac{dN}{d h_{\rm s}^2} \exp\left\{-\int^\infty_{\frac{h_{\rm s}^2}{h_{\rm s,peak}^2}} \frac{dN}{dx} dx \right\}.
\label{eq:p_h_cw}
\end{equation}
The derivation is provided in Appendix~\ref{app:p_hmax}. 
Whereas the PDF of $h_{\rm t-1}$ is calculated as Equation~\ref{eq:p_h_tot}, except the integral presumed by the $\mathcal{F}$ operation is performed only up to $h_{\rm cw}$, or, in practice, $x_{\rm cw}$. 
Whereas the conditional PDF of $h_{\rm t-1}$ given the brightest-source strain $h_{\rm cw}$ is obtained from the same inverse-Fourier construction as $p(h_{\rm t}^2)$, but with the source intensity truncated at $x_{\rm cw}=h_{\rm cw}^2/h_{\rm s,peak}^2$, so that only sources with $x < x_{\rm cw}$ contribute.

\begin{figure}
  \centering
  \includegraphics[width=\columnwidth]{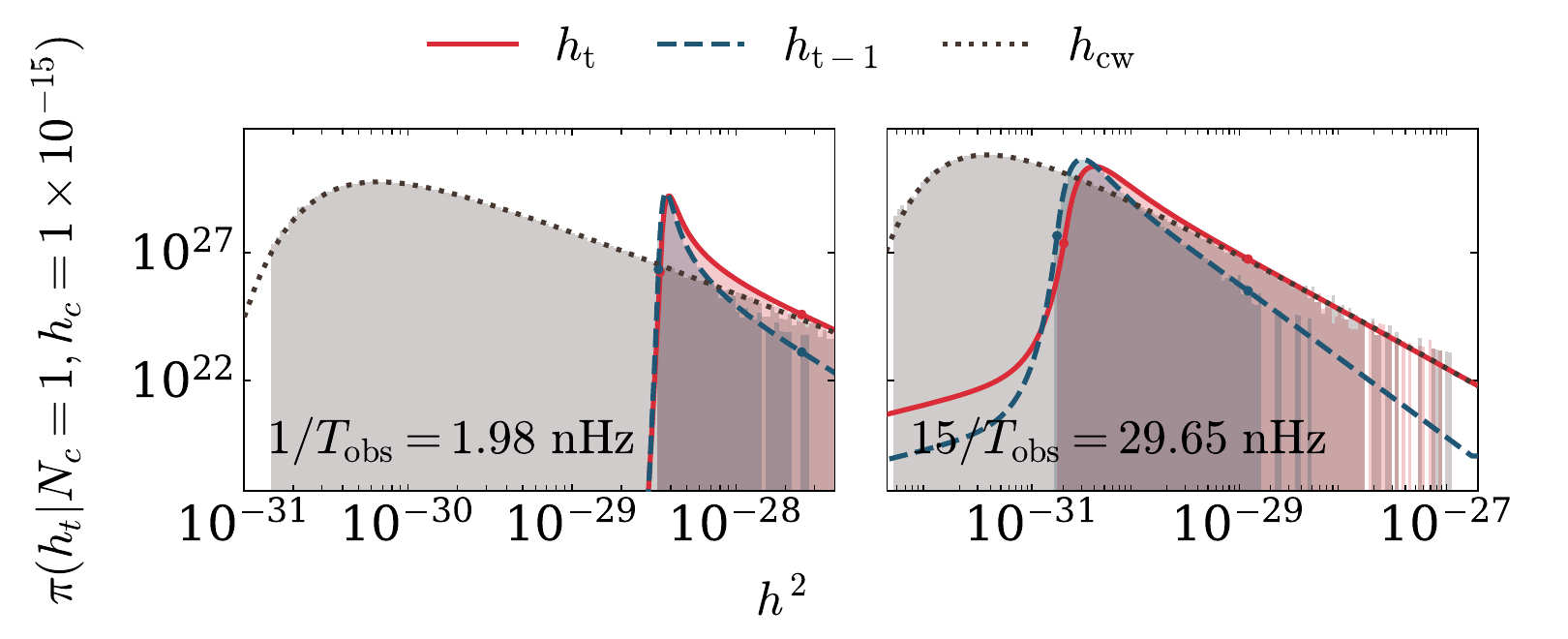}  

  \caption{An illustration of the PDF of the total characteristic strain squared for SMBHBs, $h_{\rm t}^2$, the characteristic strain of the brightest source $h_{\rm cw}^2$, and the characteristic strain of all sources except the brightest, $h_{\rm t-1}^2$. The left panel and the right panel correspond to approximately the inverse of 16~yr and the inverse of yr. }
  \label{fig:pdfs}
\end{figure}

We demonstrate PDFs for $h_{\rm t}^2$, $h_{\rm cw}^2$, and $h_{\rm t-1}^2$ in Figure~\ref{fig:pdfs}. 
The illustration is for a PTA observation of $T_{\rm obs}=16~{\rm yr}$, assuming fiducial $(N_{\rm c},h_{\rm c})=(1,10^{-15})$. 
At the lowest frequency of $T_{\rm obs}^{-1}$, $h_{\rm cw}^2$ is unlikely to exceed $h_{\rm t-1}^2$ because of the high abundance of contributing SMBHBs to the latter. 
At a higher frequency $15 T_{\rm obs}^{-1}$, it is more likely that the brightest source strain will exceed that of the remaining sources. 
It is also visible in Figure~\ref{fig:pdfs} that the power-law tail of the total characteristic strain PDF is determined by the brightest source. 
The PDF between the two filled circles is calculated directly, whereas the PDF outside of the two filled circles is calculated by extrapolation to ensure a tractable computational cost for posterior sampling. 
The PDF region calculated directly can be extended by increasing the grid on which the Fourier transforms and inverse Fourier transforms intrinsic to Equation~\ref{eq:p_h_tot} are calculated. 

From the Bayesian standpoint, our model of $\pi(h^2|N_{\rm c},h_{\rm c})$ derived from the first principles is the correct, optimal model for a superposition of all GW-driven SMBHB inspirals in circular orbits. 
The summary of our hierarchical astrophysical model is provided in Table~\ref{tab:joint_gwb_cw_model}. 
It allows the flexible treatment of resolved and unresolved SMBHB sources, as well as marginalization over unresolved sources. 
The unresolved component is treated as the isotropic GWB, a stochastic strain time series governed by Hellings-Downs correlations representing an average of the PTA's geometrical antenna response to GWs of General Relativity. 
Resolved components are treated as directional, continuous gravitational waves (CWs), with an appropriate antenna response to point sources of GWs. 
Based on a specific parametrization, the model can be used for three types of CW searches:
\begin{enumerate}
    \item $h_{\rm cw}(f_i)$: the search for one CW across all GW frequencies;
    \item $h_{{\rm cw},i}(f_i)$: the search for CW at every GW frequency; 
    \item $h_{\rm cw}(f_0)$: the search for one CW at a given frequency $f_0$ (targeted).
\end{enumerate}
The strain amplitude parameter $h_0$ used in CW searches, applied to the brightest source, is related to the characteristic strain of a CW, $h_{\rm cw}$, as
\begin{equation}
  h_0 = 2 \frac{(G \mathcal{M})^{5/3}}{D_{\rm L} c^4} (\pi f)^{2/3} = 2 h_{\rm cw} \sqrt{\frac{5}{32} \frac{\Delta f}{f}},
  \label{eq:cw_h0_hcw}
\end{equation}
where $\mathcal{M}$ is the SMBHB chirp mass, $D_{\rm L}$ is the luminosity distance, $G$ is the Newton's constant, and $c$ is a speed of light in a vacuum. 
More details are provided in Appendix~\ref{app:strain_units}.
Joint parameters $(N_{\rm c},h_{\rm c})$, which can be recast in terms of the SMBHB number density and mass scale~\cite{Sato-PolitoZaldarriaga2025}, enable an exchange of information between the inference of the GWB and the inference of CWs. 
This way, astrophysical constraints from unresolved SMBHBs influence constraints from resolved SMBHBs, and vice versa. 
When modelling brightest sources as individually resolvable CWs, our prior on $h_{\rm cw}$ narrows down on the parameter region where such signals can be identified, allowing us to reduce false positives and the Occam's penalty. 
The latter is expected to lead to stronger Bayesian evidence. 

\begin{table}[!htb]
\centering
\caption{Parameters of a joint model describing the GWB as the contribution from unresolved SMBHBs and CWs as brightest SMBHB sources. Hyperparameters $(N_{\rm c}, h_{\rm c})$ are specified at $f_{\rm ref}={\rm yr}^{-1}$ throughout this work. At frequencies where $h_{\rm cw}$ is introduced, the remaining sources are modelled as $h_{\rm t-1}$. At other frequencies, the strain is modelled as $h_{\rm t}$.}
\renewcommand{\arraystretch}{1.5}
\setlength{\tabcolsep}{8pt}
\begin{tabular}{|>{\centering\arraybackslash}p{0.44\columnwidth}|
                  >{\centering\arraybackslash}p{0.44\columnwidth}|}
\hline

\multicolumn{2}{|c|}{
  \begin{tabular}{@{}c@{}}
    \textbf{Astrophysical hyperparameters} \\[2pt]
    $(N_{\rm c},\; h_{{\rm c}})$ 
  \end{tabular}
} \\
\hline
\multicolumn{2}{|c|}{
  \begin{tabular}{@{}c@{}}
    Astrophysical hyperpriors \\[2pt]
    $\pi(N_{\rm c}),\; \pi(h_{\rm c})$ 
  \end{tabular}
} \\

\hline
\hline

\textbf{GWB} & \textbf{CW} \\
\hline
\multicolumn{2}{|c|}{
  Parameters at frequencies $f_i$, $i=(1,\ldots,N_f)$
} \\
\hline
$h_{{\rm t},i}(f_i)$ & $h_{{\rm cw},i}(f_i)$ \\ 
\hline
\multicolumn{2}{|c|}{
  Parametrized signal-specific priors
} \\
\hline
$\pi(\{h_{{\rm t},i}\} \mid N_{\rm c}, h_{{\rm c}})$ &
$\pi(\{h_{{\rm cw},i}\} \mid N_{\rm c}, h_{{\rm c}})$ \\ 

\hline
\hline

\multicolumn{2}{|c|}{\textbf{A joint hierarchical GWB+CW model}} \\
\multicolumn{2}{|c|}{
  $\pi(\{h_{{\rm t},i}|h_{{\rm t-1},i}\},\; h_{{\rm cw},i} \mid N_{\rm c}, h_{{\rm c}})$ 
} \\ 

\hline
\end{tabular}
\label{tab:joint_gwb_cw_model}
\end{table}

\section{\label{sec:methods}Methodology}


PTA data comprises the measured pulse times of arrival $t_{\rm a}$ for a set of millisecond pulsars, as well as the timing model with best-fit parameters specific to every pulsar~\cite{NG15_data}. 
The timing model describes the initial deterministic time-series predictions $t_{\rm p}$ of pulsar-specific contributions to TOAs such that the timing residuals, $\delta t=t_{\rm a}-t_{\rm p}$, are nearly flat. 
The timing model captures the contributions of pulsar spin frequency and its derivatives, astrometric and binary parameters, propagation effects in the interstellar medium, as well as telescope-specific and instrumental effects. 
The likelihood of the data $\bm{t}_{\rm a}$ is a multivariate normal distribution~\cite{vanHaasterenLevin2009}. 
It is marginalized analytically over the reduced-rank coefficients $\bm{b}$ of the time series $\bm{Tb}$ of the timing model, time-correlated stochastic processes, and epoch-correlated pulse ``jitter'', resulting in a form~\cite{NG9_GWB} 
\begin{equation}
\mathcal{L}(\bm{t}_{\rm a} | \bm{\theta}) = 
\frac{
\exp{
-\frac{1}{2} (\bm{\delta t} - \bm{s})^\intercal \bm{C}^{-1} (\bm{\delta t} - \bm{s}) 
}
}{\sqrt{ \det(2\pi \bm{C}) }},
\label{eq:likelihood}
\end{equation}
where $\bm{\theta}$ are model parameters, $\bm{C}=\bm{N}+\bm{TBT}^\intercal$ is the covariance matrix, $\bm{N}$ is the diagonal $\bm{t}_{\rm a}$-$\bm{t}_{\rm a}$ covariance matrix of the temporally-uncorrelated ``white'' noise, and $\bm{B}$ is the $\bm{b}$-$\bm{b}$ covariance matrix. 
Temporally-correlated ``red'' noise, including the GWB, has a reduced-rank representation in the Fourier domain, such that its power spectrum and inter-pulsar correlations are modeled in $\bm{B}$. 
Our methodology does not involve any modifications of the approach above, which is standard. 
Compared to standard GWB and CW searches, we only replace priors $\pi(h_{{\rm t},i})$ and $\pi(h_{{\rm cw},i})$, disconnected from each other and from the astrophysical SMBHB-specific Poisson process, with our hierarchical model $\pi(h_{{\rm t},i},\; h_{{\rm cw},i} \mid N_{\rm c}, h_{{\rm c}})$. 

The methodology outlined above is standard for PTAs, except for introducing our hierarchical model, but it is still a step ahead from the application of the original background-only model $\pi(h_{\rm t}|N_{\rm c},h_{\rm c})$ on the NANOGrav 15-year data of Ref.~\cite{Sato-PolitoZaldarriaga2025}.
There, the authors have introduced an ad-hoc factorized likelihood marginalized over $h_\text{t}(f_i)$, as shown by Equation 38 in Ref.~\cite{Sato-PolitoZaldarriaga2025}:
\begin{equation}
    \mathcal{L}(\bm{t}_{\rm a}|\Lambda) = \prod_{i=1}^{N_f} \int \frac{\mathcal{P}(h_\text{t}(f_i)|\bm{t}_{\rm a})}{\pi(h_\text{t}(f_i))} \pi(h_\text{t}^2|\Lambda) dh_\text{t}(f_i),
\end{equation}
where $\mathcal{P}(h_\text{t}(f_i)|\delta t)$ are the NANOGrav posteriors for the free-spectral model of characteristic strain. 
While this allows to loosely fit $h_{\rm t}$ to $\pi(h_{\rm t}|N_{\rm c},h_{\rm c})$, it is generally not guaranteed to yield a correct result because the full PTA likelihood is not reduced to the product mini-likelihoods containing the GWB signal only at a specific frequency bin due to a complex structure of $C$ which entangles contributions from individual frequency bins during the inversion. 
In contrast, our current approach uses the full PTA likelihood and enables a global fit to all signal and noise parameters, ensuring a complete and correct treatment of parameter estimation degeneracies.

\section{\label{sec:phenom}Phenomenology}

In this Section, we study the phenomenology of our model using simple simulated data. 
The data contains 20 pulsars timed at 100~ns precision over the period of $15.06$ years with a regular cadence of $289.47$ days. 
Thus, the data contains only $20$ pulse arrival times per pulsar. 
The pulsars are distributed isotropically over the sky, and they are described by the basic timing model, including only the linear and quadratic terms corresponding to the pulsar spin frequency and its first derivative, respectively. 
We do not simulate time-correlated noise other than the GWB to show the behaviour of our hierarchical model more clearly. 
We also neglect Hellings-Downs correlations to save on the computational cost. 
Intrinsically, this choice leads to modelling the GWB as a stochastic process carrying the expected power spectrum without consideration of the projection of the signal to pulsar pairs.
Practically, the only difference we would get by including these correlations is slightly more stringent constraints on model parameters. 

\subsection{\label{sec:phenom:gwb}GWB}

\begin{figure*}
  \centering
  \includegraphics[width=\textwidth]{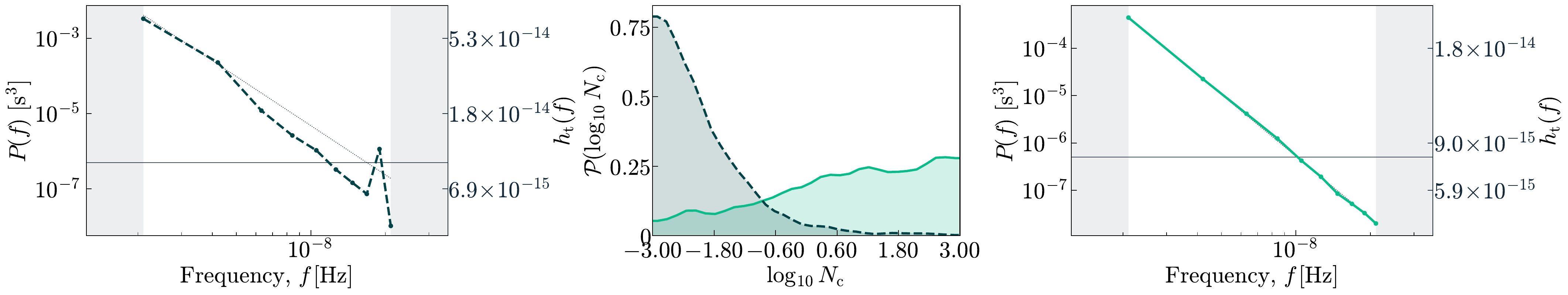} 

  \caption{
  Characteristic number of SMBHB sources, $N_{\rm c}$, as the detection statistic for the astrophysical origin of the GWB. 
  The left panel and the right panel show power spectra of the highly non-Gaussian GWB (dashed line, $N_\mathrm{c}=10^{-2.97}$) and the highly Gaussian GWB (solid line, $N_\mathrm{c}=10^{2.91}$), respectively. 
  The dotted line corresponds to rescaling $h_{\rm c}$ for these GWBs as $h_{\rm t} \propto f^{-2/3}$.
  The horizontal line shows the white noise level of the simulated PTA data with these GWBs. 
  The ticks on the right-side vertical axis show matching values of $h_{\rm t}$ for the ticks on the left-side vertical axis for the same frequency as the plotted $P(f)$.
  The middle panel shows posteriors for $N_{\rm c}$ for simulated PTA data containing these GWBs, in consistent colors and line styles. 
  By excluding the highest value of $N_{\rm c}=10^3$, we uncover the Poisson process in the GWB from a finite number of SMBHB sources. 
  } 
  \label{fig:psd_nongauss}
\end{figure*}

We start with simulating only the GWB component of our model, treating all brightest sources as unresolved. 
We only use a subset of $10$ pulsars for this simulation, for simplicity. 
We assume a fixed $h_{\rm c}=2 \times 10^{-14}$, a Uniform (hyper-)prior distribution of $\pi(\log_{10}N_{\rm c})=\mathcal{U}(-3,3)$\footnote{This model is fiducial and uninformative. An astrophysical model can also be developed, \textit{e.g.}, based on the SMBHB population synthesis, resulting in yet more optimal inference.}, and we model $h_{\rm t}$ at $10$ frequency bins. 
Drawing 500 Monte-Carlo samples from our hierarchical model, we show two specific realizations of $h_{\rm t}$, as well as the results of parameter estimation on simulated data based on these realizations, in Figure~\ref{fig:psd_nongauss}. 
The left and the right panels show power spectra of specific realizations in the units of the power spectral density of timing residuals~$[{\rm s}^3]$ and in the units of characteristic strain:
\begin{equation}
    P(f) = \frac{h^2_{\rm t}(f)}{12 \pi^2 f^3}.
\end{equation}
A realization with the low $N_{\rm c}$ shown on the left panel produces visible deviations from the power-law characteristic strain, a mean of the RMS strain PDF, shown as the dotted line. 
A realization with high $N_{\rm c}$ shown on the right panel matches the power-law approximation. 
The middle panel shows the marginalized posterior of $\log_{10}N_{\rm c}$ for the two realizations mentioned above. 

The limiting case of a reduction of the strain power spectrum to the power law, shown in the right panel of Figure~\ref{fig:psd_nongauss}, corresponds to the Gaussian stochastic GWB. 
A more general case corresponding to the left panel in Figure~\ref{fig:psd_nongauss} features fluctuations in the power spectrum because the PDFs intrinsic to our model (shown in Figure~\ref{fig:pdfs}) are non-Gaussian.
This adds a non-Gaussian component to the otherwise Gaussian stochastic process. 
In other words, the time-domain covariance of the GWB can not be described by the Gaussian likelihood function alone. 
Therefore, we refer to these observable fluctuations in the power spectrum of the GWB as non-Gaussianity. 
This is not to be confused with non-Gaussian inflationary perturbations attributed to some of the GWBs from the early Universe~\cite{NG15_NEWPHYS}. 


Figure~\ref{fig:psd_nongauss} hints at the possibility of using $N_{\rm c} = 1000$ as the null hypothesis for our astrophysical GWB model. 
At $N_{\rm c} \geq 1000$, our model of a superposition of SMBHB point sources is indistinguishable from a Gaussian GWB.
The latter represents the limit of an infinite number of sources. 
A measurement of $N_\mathrm{c}$ that excludes this value can be taken as the clear signature of the SMBHB-driven Poisson process in the gravitational wave background, assuming the consistency of the average inferred power spectrum with the mean of the predicted PDF of the RMS strain has been verified as well. 
This also represents an observation of the non-Gaussianity of the GWB. 
Although the span of our prior range is ultimately influenced and limited by the chosen resolution of the Fourier grid on which RMS strain PDFs are computed, as described in Section~\ref{sec:model}, the edges of the prior correspond to the highly-Gaussian GWB and highly-non-Gaussian GWB cases. 
Treating the threshold $N_{\rm c} = 1000$ as the null hypothesis, a Savage-Dickey Bayes factor can be computed in favor of our astrophysical model. 
It is visible in Figure~\ref{fig:psd_nongauss} that the posterior marked by the dashed line in the middle panel would result in a significant Bayes factor. 
The threshold of $N_{\rm c} = 1000$ can be chosen arbitrarily, keeping in mind that lower values of the threshold could make the detection of the Poisson nature of the GWB more difficult. 
In practice, another limitation of the parameter estimation for $N_{\rm c} > 1000$ is Neal's funnel problem. 
PDFs of the RMS strain become very narrow, which makes it difficult for posterior samplers to explore them.


\subsection{\label{sec:phenom:cw}GWB with one brightest source}

\begin{figure*}
  \centering
  \includegraphics[width=\textwidth]{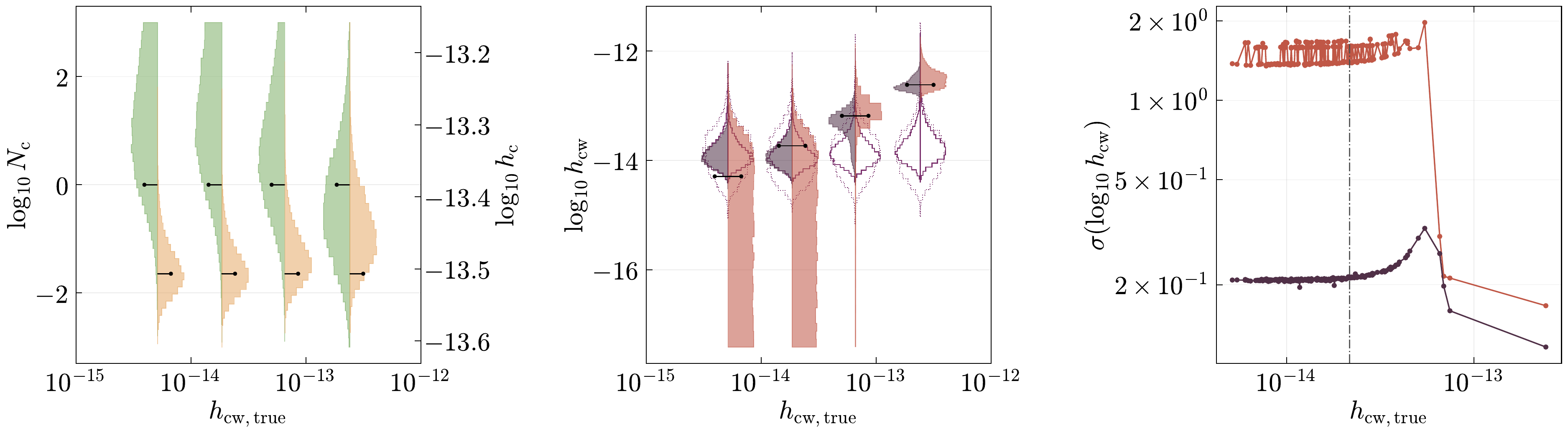} 

  \caption{The estimation of parameters of our joint model of the GWB and continuous GWs from SMBHBs as a function of the continuous GW brightness, $h_{\rm cw,true}$. 
  The left panel shows marginalized posteriors for $N_{\rm c}$ (left-facing violins) and $h_{\rm c}$ (right-facing violins). 
  Values along the horizontal axis, $h_{\rm cw,true}$, correspond to $4$ approximately equally-spaced values of $250$ Monte-Carlo samples from $\pi(h_{\rm cw}|N_{\rm c,true}=1, h_{\rm c,true}=2 \times 10^{-14})$. 
  Black lines show simulated values $(N_{\rm c,true}, h_{\rm c,true})$. 
  Each posterior is based on the data simulated with different $h_{\rm cw,true}$ and the same $h_{\rm t,true}$. 
  The middle panel shows marginalized posteriors for $h_{\rm cw}$ with our joint model for GWB and CWs (left-facing blue histograms) and with the standard agnostic model (right-facing red histograms). 
  The dotted hollow contours in the middle panel show the prior-predictive distribution, and the solid hollow contours show the posterior-predictive values. 
  The right panel shows the $1\sigma$ credible level for $h_{\rm cw}$ as a function of $h_{\rm cw,true}$ for all $250$ simulation realizations, including $4$~realizations presented in the other panels. 
  The vertical dash-dotted line corresponds approximately to the white noise level in the simulation. 
  The density of points follows Figure~\ref{fig:pdfs} and demonstrates that the rightmost $h_{\rm cw, true}$ across all panels is the rare but expected outlier. 
  } 
  \label{fig:cwh_vs_cw_vs_h}
\end{figure*}

Here, we simulate the stochastic GWB component of our model alongside the brightest source at $2\times10^{-8}$ Hz, close to the 10th Fourier frequency bin.
We simulate one realization of $h_{\rm t}$ at all frequencies except that of the brightest source with a fixed $h_{\rm c}=3\times10^{-14}$ and $N_{\rm c}=1$.
At the frequency of the brightest source, we simulate $250$ realizations of $h_{\rm cw}$, treating the brightest source as a CW, and keeping $h_{\rm t-1}$ fixed. 
This allows us to inspect the effect of SMBHB source brightness without a distortion from random fluctuations of other parameters. 
We still assume a uniform (hyper-)prior distribution of $\pi(\log_{10}N_{\rm c})=\mathcal{U}(-3,3)$ and $\pi(\log_{10}h^2_{\rm c})=\mathcal{U}(-28,-26)$ to show full parameter estimation capabilities.
The results of parameter estimation on the simulated data are shown in Figure~\ref{fig:cwh_vs_cw_vs_h}. 

The left panel of Figure~\ref{fig:cwh_vs_cw_vs_h} shows marginalized posteriors on $\log_{10}N_{\rm c}$ and $\log_{10}h_{\rm c}$ as a function of simulated $h_{\rm cw}$ values, $h_{\rm cw,true}$, for four selected simulations including those with minimum $h_{\rm cw}$ across $250$ realizations, maximum $h_{\rm cw}$, and a few in between such that they are approximately equally-spaced. 
The posterior density is shown along the horizontal axis, unlabelled, in parallel with $h_{\rm cw}$, facing towards the left for $\log_{10}N_{\rm c}$ and facing towards the right for $\log_{10}h_{\rm c}$. 
Black lines represent the simulated values. 
The rightmost posterior of $\log_{10}N_{\rm c}$ at the highest, exceptionally rare $h_{\rm cw}$ visibly suggests the strongest evidence in favor of our astrophysical model by preferring $N_{\rm c}<1000$.
Thus, it also corresponds to the detection of the Poisson process and non-Gaussianity associated with SMBHBs. 
This is because, on average, stronger outliers in $h_{\rm cw}$ are compatible with higher $\log_{10}N_{\rm c}$. 
From the astrophysical standpoint, both unresolved $\log_{10}N_{\rm c}$ and $\log_{10}h_{\rm c}$ suggest no evidence of the gravitational wave background from SMBHBs. 
When only $\log_{10}h_{\rm c}$ is resolved and $\log_{10}N_{\rm c}$ is unresolved, it implies a degeneracy between the SMBHB number density and the mass scale~\cite{Sato-PolitoZaldarriaga2025}. 
When both parameters are resolved, the degeneracy is broken. 

The middle panel of Figure~\ref{fig:cwh_vs_cw_vs_h} shows marginalized posteriors on $\log_{10}h_{\rm cw}$ as a function of $h_{\rm cw,true}$, demonstrating the effect of our model on parameter estimation. 
The marginalized posteriors in this panel correspond to the same four posteriors used to produce the marginalized posteriors in the left panel. 
The posteriors facing left in the middle panel of Figure~\ref{fig:cwh_vs_cw_vs_h} correspond to our hierarchical model, whereas the posteriors facing right correspond to the standard CW search with a uniform prior on $h_{\rm cw}$. 
For the two simulations with the lowest $h_{\rm cw}$, right-facing posterior tails extending to $\log_{10}h_{\rm cw}=-18$ indicate that the signals are unresolved, undetected. 
For the two simulations with the highest $h_{\rm cw}$, a lack of density at $\log_{10}h_{\rm cw}=-18$ in right-facing posteriors suggests that the signals are resolved, detected. 
The tails are absent on posteriors facing left because our hierarchical model does not allow arbitrary low $h_{\rm cw}$ with our specific choice of (hyper-)priors on $(N_{\rm c},h_{\rm c})$. 
Hyper-prior-predictive distribution of $h_{\rm cw}$ corresponding to 
\begin{equation}
\int \pi(h_{\rm cw}|N_{\rm c},h_{\rm c}) \pi(N_{\rm c}) \pi(d h_{\rm c}) d N_{\rm c} d h_{\rm c}, 
\end{equation}
is shown as the dotted line. 
It shows the effective range of $h_{\rm cw}$ allowed by $\pi(N_{\rm c},h_{\rm c})$.
Unlike for the uniform prior assumed for the right-facing posteriors, our model makes it difficult to see by eye whether the CW signal is detected. 
So, the use of Bayesian odds and Bayesian evidence is necessary. 
The middle panel of Figure~\ref{fig:cwh_vs_cw_vs_h} also illustrates that the highest simulated value of $h_{\rm cw}$ is not consistent with the bulk of the hyper-prior predictive distribution due to a high number of trials (250). 
It is slightly more consistent with posterior-predictive and $(\hat{N}_{\rm c},\hat{h}_{\rm c})$-predictive distributions, which are the left-facing and right-facing in solid lines, respectively. 
Throughout this work, the hat symbol denotes the best-fit value (maximum-\textit{aposteriori}). 

The right panel of Figure~\ref{fig:cwh_vs_cw_vs_h} shows the span of the $1\sigma$ credible level for $\log_{10}h_{\rm cw}$ for all $250$ simulations, which include $4$ simulations analyzed in the other panels. 
The bottom curve corresponds to our model. 
The top curve corresponds to the uniform prior for $\log_{10}h_{\rm cw}$. 
The density of points around simulated $h_{\rm cw}$ values illustrates how likely it is to obtain these values based on our model and selected hyperparameters. 
The vertical dash-dotted line approximately corresponds to the white noise level in the units of $h_{\rm cw}$. 
At this level, the measurement uncertainty of $\log_{10}h_{\rm cw}$ decreases, suggesting that the signal is distinguished from the noise. 
Please note that more stringent constraints with our model are due to our more restrictive hierarchical prior.
While it suggests superior parameter estimation capabilities of our model, it does not automatically imply that the signal is detected more efficiently. 
It is visible, however, that the signal becomes distinguishable from noise by the data (\textit{i.e.}, recognized by the likelihood function) at $h_{\rm cw} \approx 8 \times 10^{-14}$, simultaneously with the non-hierarchical model. 
The detection capabilities of our model compared to the standard, non-hierarchical CWs model can be explored in future work.

\section{\label{sec:results}Insights from the 15-year NANOGrav data}

We perform the joint search for resolved and unresolved nanohertz-frequency GWs from GW-driven SMBHB inspirals in the simulated NANOGrav 15-year data~\cite{NG15_data} based on our new model. 
We simulate everything that is known about the NANOGrav data: pulse arrival times and their measurement uncertainties, backend-receiver combinations, white and red noise, as well as the inferred GWB power spectrum. 
We neglect the Hellings-Downs correlations because they have a negligible effect on our results. 
The results of an analysis with the real NANOGrav 15-year data are in the process of review within the International Pulsar Timing Array (IPTA~\cite{IPTA_DR1}). 
The conclusions of that analysis are in line with what we present here.

\begin{figure}
  \centering
  \includegraphics[width=\columnwidth]{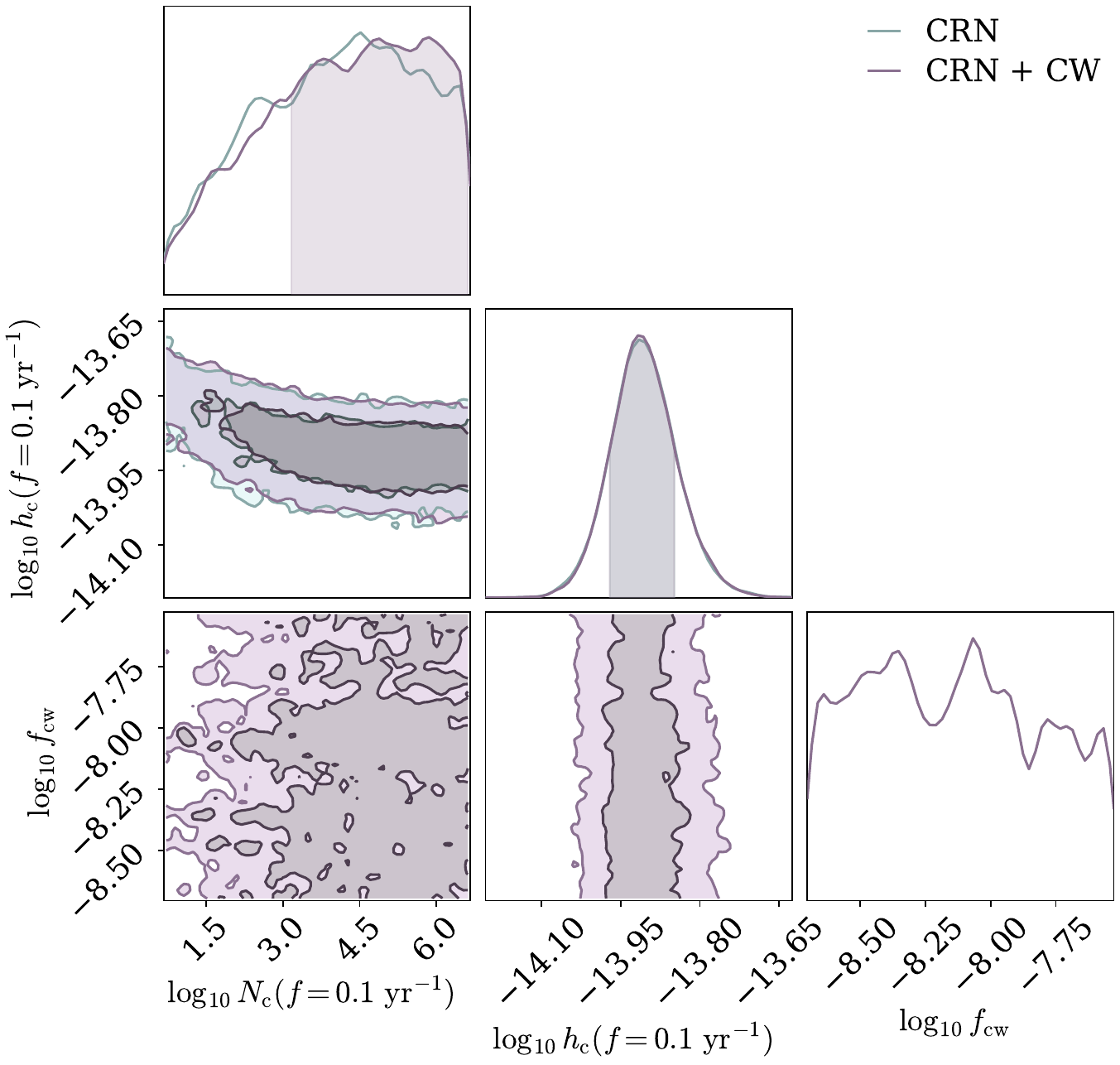}  

  \caption{Marginalized posterior for three selected parameters of our joint hierarchical model for GWs from resolved and unresolved SMBHBs with the simulated NANOGrav 15-year data: the characteristic number of sources $N_{\rm c}$ at $10~{\rm yr}^{-1}$, the characteristic strain amplitude $h_{\rm c}$ at $10~{\rm yr}^{-1}$, and the frequency $f_{\rm cw}$ of the fitted brightest source of GWs.}
  \label{fig:posterior_Nc_hc_fcw}
\end{figure}

We start off with a search for the astrophysical GWB without individually resolvable sources.
Next, we search for the GWB with the brightest SMBHB source, which may be individually resolvable as a CW. 
In both cases, we search for GW signals with frequencies spanning from $T_{\rm obs}^{-1}=1.98$~nHz to $15 T_{\rm obs}^{-1}=29.65$~nHz. 
Beyond these frequencies, the data is not expected to be sensitive to astrophysical GWs from SMBHBs. 
The results of parameter estimation for key parameters of our model --- the characteristic number of SMBHB sources $N_{\rm c}$, the characteristic strain amplitude $h_{\rm c}$, both at $f={\rm yr}^{-1}$, and the frequency of the brightest source $f_{\rm cw}$ --- are shown in Figure~\ref{fig:posterior_Nc_hc_fcw}. 
The results are shown for both the GWB-only model, which assumes no individually resolved SMBHBs as CWs and an isotropic GW distribution in the sky featuring Hellings-Downs correlations, as well as for the model which assumes one additional brightest source that may be resolved at a fitted frequency $f_{\rm cw}$. 

The most important observation from Figure~\ref{fig:posterior_Nc_hc_fcw} is that there is insufficient evidence for the astrophysical origin of the GWB in our simulated NANOGrav 15-year data. 
Namely, strong consistency of the inferred $N_{\rm c}$ with the value of $N_{\rm c}=1000$ of our astrophysical null hypothesis means that we do not find evidence for the SMBHB-related Poisson process in the GWB. 
The parameter estimation of $N_{\rm c}$ is robust to the assumption about the presence of individually resolvable SMBHB sources. 

\begin{figure}
  \centering
  \includegraphics[width=\columnwidth]{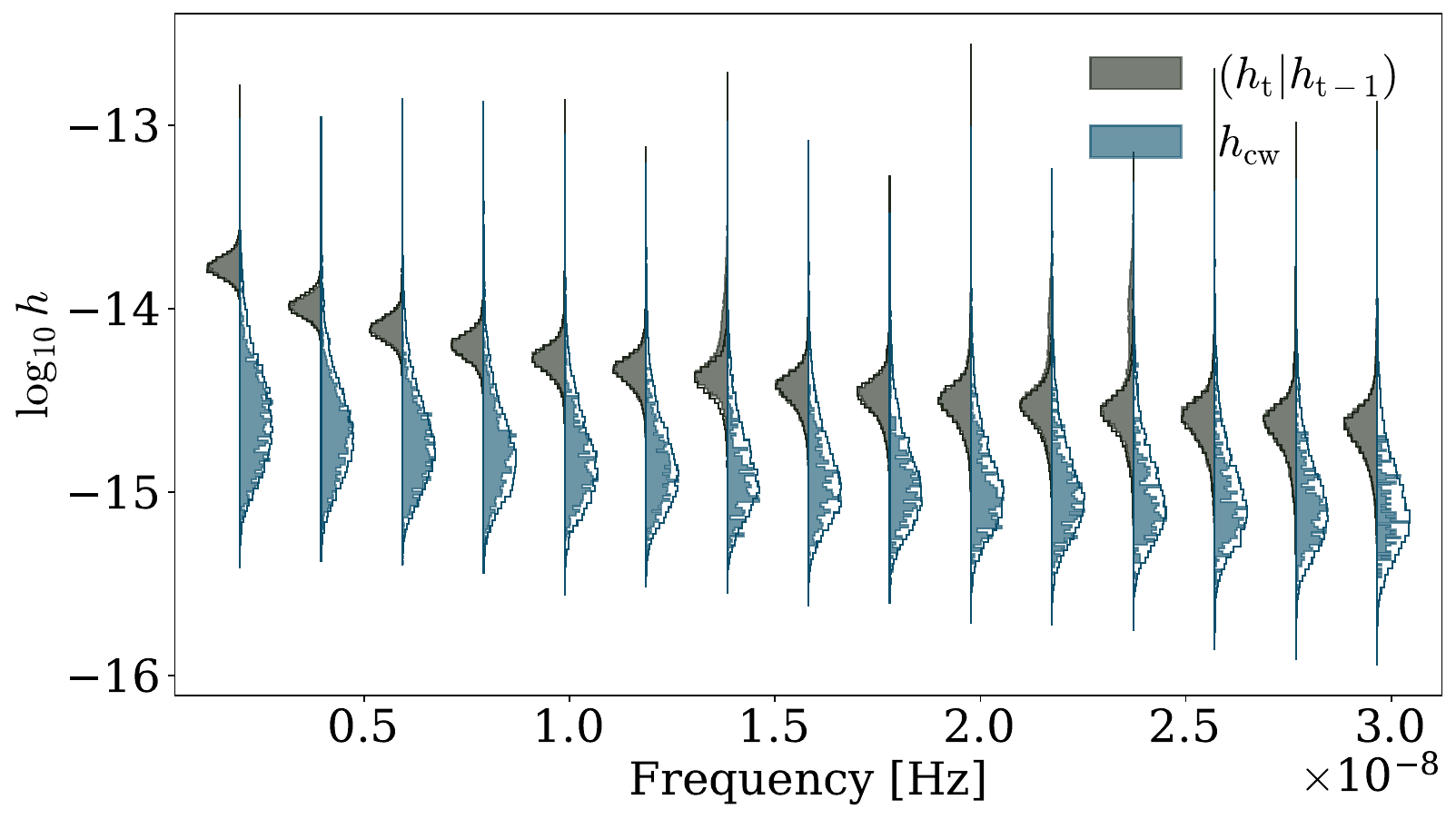}  

  \caption{Marginalized posteriors (filled) and posterior-predictive samples (hollow) for the characteristic strain of the unresolved gravitational wave (GW) sources (left) and the resolved sources (right) for every analyzed GW frequency bin. 
  Towards high frequencies, the total characteristic strain is increasingly dominated by the brightest sources, which may become resolvable, but the scaling of the PTA white noise with characteristic strain as $\propto f^{3/2}$ diminishes the sensitivity. 
  }
  \label{fig:hcw_ht_p_pp}
\end{figure}

The second important observation from Figure~\ref{fig:posterior_Nc_hc_fcw} is that there is no evidence for a resolvable CW in our simulated NANOGrav 15-year data~\cite{NG15_GWB,NG15_SMBHB}. 
It is visible from the posterior support for $\log_{10}f_{\rm cw}$ across the whole range of the prior. 
A lack of individually-resolvable GW signals from SMBHBs is also visible when comparing posteriors on the characteristic strain of the GWB ($\{h_{\rm t},h_{\rm t-1}\}$) and the expected brightest sources ($h_{\rm cw}$) with the posterior-predictive values generated from $\pi(h_{\rm t},h_{\rm cw}|\{N_{\rm c},h_{\rm c}\}_{\mathcal{P}})$. 
We show the samples from these distributions in Figure~\ref{fig:hcw_ht_p_pp}. 
Matching densities for $h_{\rm cw}$ suggest there are no exceptionally bright CWs in the data. 

The remaining subsections are organized as follows. In Subsection~\ref{sec:results:limits}, we report the most stringent limits of the CW strain and demonstrate the tension of these limits with several predicted SMBHB candidates from Active Galactic Nuclei (AGN). 
In Subsection~\ref{sec:results:snr}, we predict the detection of the CW signal based on our astrophysical model in the existing 15-year NANOGrav data and the future 20-year NANOGrav data.
In Section~\ref{sec:astro}, we provide the astrophysical interpretation of our results.

\subsection{\label{sec:results:limits} Astrophysical limits on the strain amplitude of CW from SMBHBs}

\begin{figure}
  \centering
  \includegraphics[width=\columnwidth]{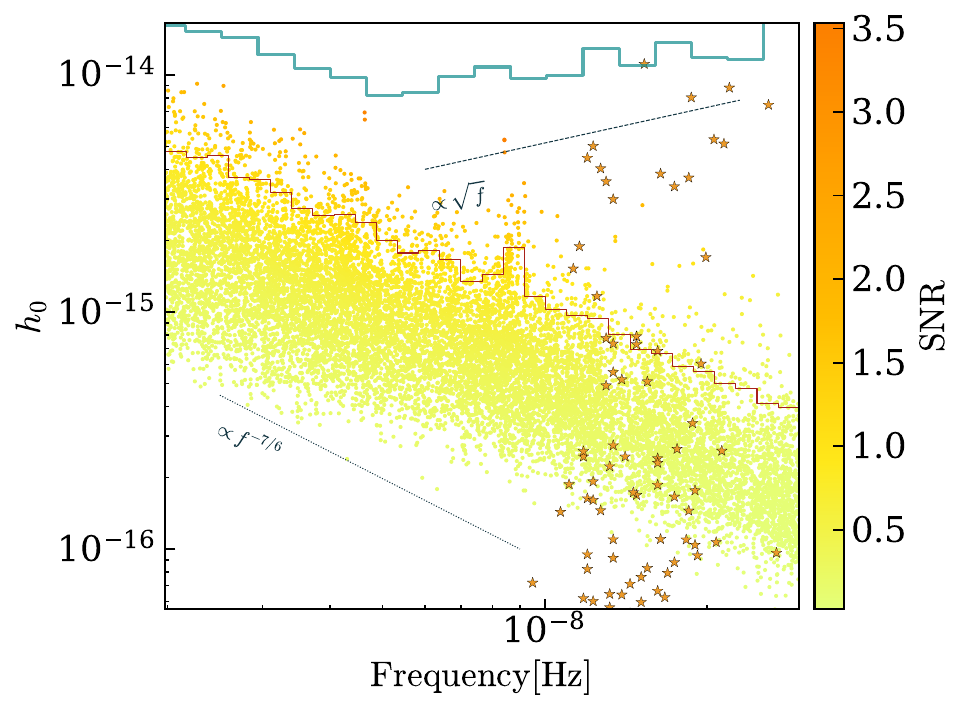} 

  \caption{
  Constraints on the strain amplitude $h_0$ of CWs from brightest SMBHBs across the sky with the simulated NANOGrav 15-year data.  
  Dots show posterior samples from our joint search for CWs and the GWB. 
  Solid red binned lines show the upper limits at 95\% credibility based on our posterior. 
  In contrast, the cyan binned line shows the limit reported by NANOGrav~\cite{NG15_SMBHB} based on placing an astrophysically-agnostic prior on $h_0$. 
  Stars show SMBHB candidates for the targeted CW search with the NANOGrav 15-year data~\cite{NG15_targeted}. 
  }
  \label{fig:limits}
\end{figure}

Based on our new methodology, we place limits on the CW strain amplitude $h_0$ from the brightest SMBHB sources with the simulated NANOGrav 15-year data. 
We report the posterior $\mathcal{P}(h_0,f_0)$ of our search in Figure~\ref{fig:limits}. 
Upper limits on the CW strain at 95\% credibility obtained with our hierarchical model are shown in red. 
These limits are informed by the fitness of the data to all GWs, including the resolved GWB. 
The limits on CW strain obtained by Ref.~\cite{NG15_SMBHB} without any assumptions about the astrophysics of SMBHBs are shown in cyan. 
Our improved limits show that more of the SMBHB candidates from AGN observations~\cite{GrahamDjorgovski2015,O'NeillKiehlmann2022,delaParraKiehlmann2025,SudouIguchi2003}, shown as stars, are in tension with the NANOGrav observations. 

Unlike in Figure~\ref{fig:hcw_ht_p_pp}, which shows characteristic strain, posterior samples in Figure~\ref{fig:limits} decline more rapidly with frequency due to a factor of $\Delta f / f$ from Equation~\ref{eq:cw_h0_hcw}. 
At high frequencies, the all-sky limit from NANOGrav is dominated by white noise, which scales as $\propto f^{3/2}$ in characteristic strain and $\propto \sqrt{f}$ in strain amplitude $h_0$. 
Whereas our limit is driven by nearly-$f^{-2/3}$ in characteristic strain and thus $\propto f^{-7/6}$ in $h_0$.

The color of posterior samples in Figure~\ref{fig:limits} shows the posterior on the signal-to-noise ratio parameter ${\rm SNR}$, which is our expectation of the inferred CW ${\rm SNR}$ in the NANOGrav 15-year data if there was a CW signal with the amplitude given by the posterior sample. 
Calculating ${\rm SNR}$ enables astrophysically-agnostic, observational insights on CWs from the posterior, which is otherwise significantly driven by our astrophysical SMBHB model and the constraints from the GWB. 
The SNR parameter is calculated as 
\begin{equation}
  {\rm SNR} = \frac{h_0}{\sqrt{S_{\rm eff}(f)}} \sqrt{T_{\rm obs}},
  \label{eq:snr}
\end{equation}
where $S_{\rm eff}(f)$ is the strain power spectral density of noise calculated for NANOGrav pulsars based on the methodology from Ref.~\cite{HazbounRomano2019}, posterior samples of $h_0$, the best-fit white noise parameters, and the posterior samples of red noise. 
Finding ${\rm SNR}<4$ suggests that the detection of a CW signal in the NANOGrav 15-year data is unlikely, consistent with the recent results~\cite{NG15_SMBHB,NG15_targeted}. 


\subsection{\label{sec:results:snr} Posterior-predictive SNR and CW detection probability}

In this Subsection, we calculate the detection probability of the CW signal from the brightest SMBHB sources in the all-sky search for the NANOGrav 15-year data, as well as the projected CW detection probability for the 20-year data. 

The calculation is performed as follows. 
Based on our posterior samples of $S_{\rm eff}(f)$, and based on posterior-predictive samples of $h_0$, we calculate the posterior-predictive SNR (${\rm SNR}_{\rm p}$). 
The calculation is as in Equation~\ref{eq:snr}, except for posterior samples on $h_0$ substituted with posterior-predictive samples. 
This approach is recommended for the analysis of real data in order not to contaminate the prediction with any existing CW outliers that may exist in the data. 
Extrapolating TOA and TOA errors of pulsars based on the last year of their observations $5$ years into the future, we also obtain an approximation of the $S_{\rm eff}(f)$ for the expected NANOGrav 20-year data. 
The increase in frequency resolution for future data is neglected. 
The detection probability is defined as the fraction of ${\rm SNR}_{\rm p}$ samples exceeding the threshold of $5$. 

\begin{figure*}
  \centering
  \includegraphics[width=\textwidth]{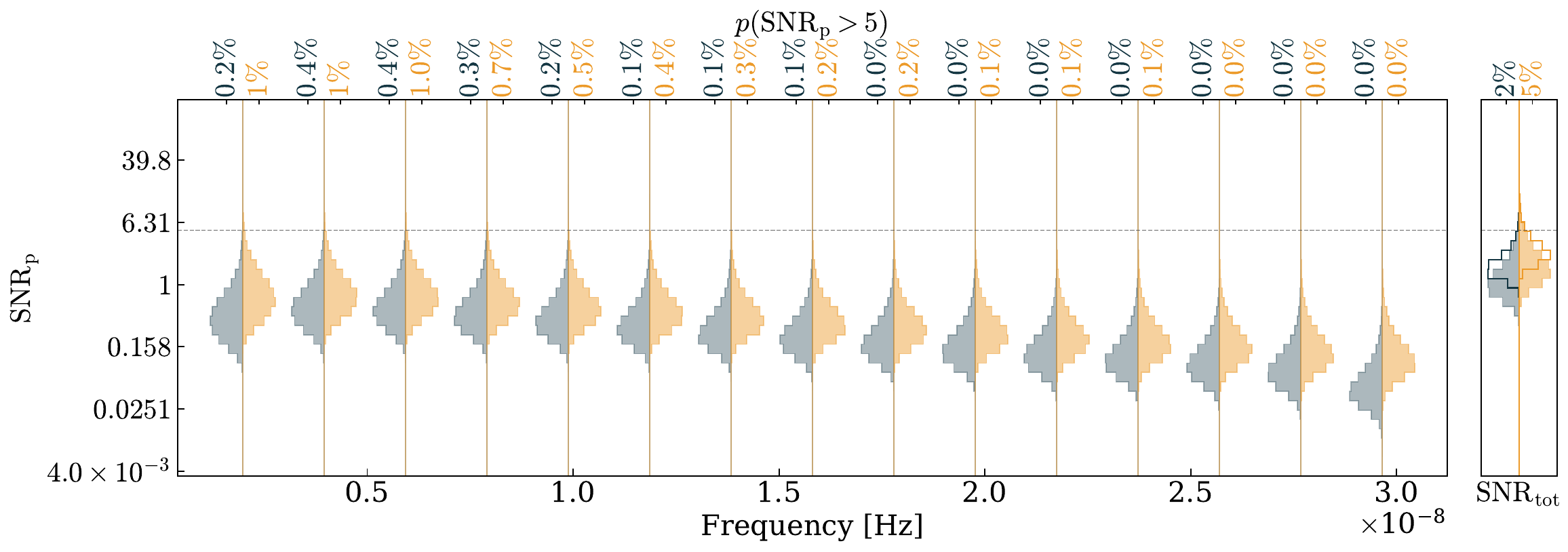}

  \caption{Posterior-predictive signal-to-noise ratio ${\rm SNR}_{\rm p}$ and detection probability for continuous gravitational waves (CWs) in the NANOGrav data in the frequency range from $T_{\rm obs}^{-1}$ to $15~T_{\rm obs}^{-1}$. 
  Here, $T_{\rm obs}=16.02~{\rm yr}$ corresponding to the NANOGrav 15-year data.
  The left-facing histograms show ${\rm SNR}_{\rm p}$ samples for $15$ frequency bins corresponding to the 15-year data, whereas the right-facing histograms show samples for the 20-year data corresponding to five extra years of observation. 
  The right panel shows samples of maximum ${\rm SNR}_{\rm p}$ across frequencies (filled), as well as the root-mean-squared across frequencies (hollow). 
  The dashed horizontal line corresponds to the fiducial detection threshold ${\rm SNR}_{\rm p}=5$.
  The top horizontal axes show the fraction of ${\rm SNR}_{\rm p}$ samples exceeding this threshold, quantifying the detection probability. 
  }
  \label{fig:psd_snr_predicted}
\end{figure*}

We show ${\rm SNR}_{\rm p}$ for the 15-year NANOGrav data (left-facing histograms), as well as for the projected 20-year NANOGrav data (right-facing histograms), in Figure~\ref{fig:psd_snr_predicted}. 
Per-frequency values are supplemented with a total SNR on the right-side summary panel. 
The primary method for calculating the total SNR is based on plotting the maximum SNR value from each frequency bin for each sample. 
It corresponds to the all-sky all-frequency search for a CW signal, and it corresponds to filled histograms. 
The secondary method for calculating the total SNR is based on the RMS sum of per-frequency SNRs. 
It may correspond to the hypothetical search for CWs at all frequencies simultaneously, which we will explore in future work. 
Such SNR samples are shown as hollow histograms. 
The detection probability is shown along the top horizontal axis. 
The right panel shows that we estimate the probability of a CW detection to be only $2\%$ in the NANOGrav 15-year data and $5\%$ in the expected 20-year data. 
Thus, we find a significantly lower probability of detecting a CW with the NANOGrav data than in Ref.~\cite{GardinerBecsy2025}. 
However, our detection probabilities are consistent with Figure~2 in Ref.~\cite{RosadoSesana2015}.
This figure indicates that the IPTA at 15 years of observation is associated with about 5\% CW detection probability and about 70\% GWB detection probability. 

Even though a confident CW detection is unlikely for the current and projected NANOGrav data, there is a non-negligible probability of finding a low-significance CW outlier. 
In Table~\ref{tab:cw_snr_detection_probabilities}, we report probabilities for finding CW signals exceeding SNRs ranging between 2 and 5. 
We report the probability of exceeding the SNR threshold in both the all-sky all-frequency search for one signal ($p_{\rm max}$), as well as when searching for a superposition of signals at all frequencies ($p_{\rm rms}$). 
Notably, the probability of observing emerging evidence of the CW signal at the level of SNR=2 in the expected 20-year NANOGrav data is 39\%, whereas the probability of observing combined evidence of CWs across all frequencies is 82\% (RMS statistic). 


\begin{table}[!htb]
\centering
\caption{Posterior-predictive probabilities for finding a continuous gravitational wave (CW) in the existing NANOGrav 15-year data and the expected 20-year data exceeding a given signal-to-noise ratio (SNR) threshold.}
\renewcommand{\arraystretch}{1.35}
\setlength{\tabcolsep}{3.3pt}
\begin{tabular}{|>{\centering\arraybackslash}p{0.16\columnwidth}|>{\centering\arraybackslash}p{0.22\columnwidth}|>{\centering\arraybackslash}p{0.10\columnwidth}|>{\centering\arraybackslash}p{0.10\columnwidth}|>{\centering\arraybackslash}p{0.10\columnwidth}|>{\centering\arraybackslash}p{0.10\columnwidth}|}
\hline
\multirow{2}{*}{$T_{\rm obs}$} & \multirow{2}{*}{Statistic} & \multicolumn{4}{c|}{\textbf{SNR threshold}} \\
\cline{3-6}
 & & $2$ & $3$ & $4$ & $5$ \\
\hline
\multirow{2}{*}{$15$~yr} & $p_{\rm det,max}$ & $16\%$ & $7\%$ & $3\%$ & $2\%$ \\
\cline{2-6}
 & $p_{\rm det,rms}$ & $34\%$ & $10\%$ & $4\%$ & $2\%$ \\
\hline
\multirow{2}{*}{$20$~yr} & $p_{\rm det,max}$ & $39\%$ & $17\%$ & $9\%$ & $5\%$ \\
\cline{2-6}
 & $p_{\rm rms}$ & $82\%$ & $36\%$ & $16\%$ & $8\%$ \\
\hline
\end{tabular}
\label{tab:cw_snr_detection_probabilities}
\end{table}


\section{\label{sec:astro} Astrophysical interpretation} 

Two hyperparameters of our model, $(h_{\rm c},N_{\rm c})$, which determine GWB's amplitude and its scatter, respectively, can be recast in terms of SMBHB abundance and mass scale. 
The approach below follows Ref.~\cite{Sato-PolitoZaldarriaga2025}. 
A meaningful parametrization for the mass scale is the peak of the mass kernel $\mathcal{M}^{5/3} \phi(\mathcal{M})$, which we refer to as $M_{\rm peak}$.
The factor of $\mathcal{M}^{5/3}$ defines the peak of the kernel as the SMBHB mass with the most contribution to the GWB. 
Whereas 
\begin{equation}
    \phi(M) = \int \frac{p(\log_{10}\mathcal{M}|\log_{10}\sigma)}{\mathcal{M} \log(10)} \phi(\sigma) d\sigma 
    \label{eq:MF_scatter}
\end{equation}
is the SMBHB mass function modeled as the log-10-Normal distribution 
\begin{equation}
\begin{split}
    p(\log_{10}&\mathcal{M}|\log_{10}\sigma) = \mathcal{N}(a_{\bullet}+b_{\bullet}\log_{10}\sigma,\epsilon_0)
    \label{eq:galaxy_property_scaling}
\end{split}
\end{equation}
convolved with the velocity dispersion function 
\begin{equation}
    \phi(\sigma) = \phi_{*} \left(\frac{\sigma}{\sigma_*}\right)^{\alpha} \frac{e^{-(\sigma/\sigma_*)^{\beta}}}{\Gamma(\alpha/\beta)}  \frac{\beta}{\sigma}.
    \label{eq:sigma_function}
\end{equation}
Here, $\sigma$ is the galaxy velocity dispersion [km/s]. 
Coefficients $(a_{\bullet},b_{\bullet})$ determine the connection between a galaxy's velocity dispersion and the chirp mass of its central black hole, whereas $\epsilon_0$ controls the uncertainty of this scaling relation. 
Parameters $\phi_{*}$, $\sigma_*$, $\alpha$, and $\beta$ control the velocity dispersion function. 
The quantities outlined above are used to calculate the luminosity function $dN/dh^2_{\rm s}$ in our analysis. 
It is then convenient to define the abundance of SMBHBs as 
\begin{equation}
    \rho_{\rm BH} \propto M_{\rm peak} \phi_{*}. 
\end{equation}

\begin{figure}
  \centering
  \includegraphics[width=\columnwidth]{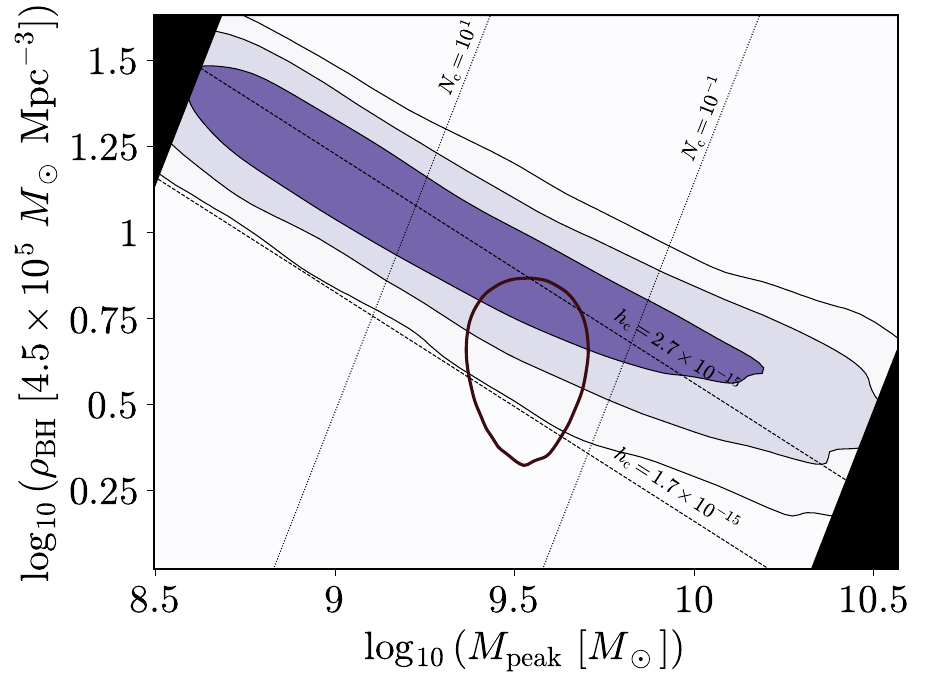}  

  \caption{Posterior on the SMBHB mass density $\rho_{\rm BH}$ and the peak of the mass kernel $M_{\rm peak}$ with the simulated NANOGrav 15-year data. 
  For comparison, the $1\sigma$ range of the Normal prior based on the properties of galaxy demographics, which was used in the EPTA fit to the Gaussian stochastic GWB described by a power-law, is shown as two dashed lines with constant $h_{\rm c}$. 
  The thick brown line shows the EPTA prior~\cite{GoncharovSardana2025a} recast in terms of our astrophysical model. 
  }
  \label{fig:rho_mpeak}
\end{figure}

To extract astrophysical insights from our observations, we recast our posterior samples of $(N_{\rm c},h_{\rm c})$ in terms of astrophysical quantities $(\rho_{\rm BH},M_{\rm peak})$ defined above. 
The result is shown in Figure~\ref{fig:rho_mpeak}.
The format is adopted from Figure~5 in Ref.~\cite{Sato-PolitoZaldarriaga2025}, except our $N_{\rm c}$ is referenced to $1~{\rm yr}^{-1}$ instead of $0.3~{\rm yr}^{-1}$. 
The purple contours show credible levels at $1$, $2$, and $3\sigma$, respectively. 
Dashed lines show fixed levels of $h_{\rm c}$ and dotted lines show fixed levels of $N_{\rm c}$. 
Black regions show edges of our prior on $N_{\rm c}$. 

For comparison with our NANOGrav results in Figure~\ref{fig:rho_mpeak}, we show the results of a more simplified fitting of the scaling of $h_{\rm c}$ with astrophysical parameters to the European Pulsar Timing Array (EPTA) Second Data Release~\cite{GoncharovSardana2025a}. 
This analysis is performed in the limit of the infinitely large $N_{\rm c}$, as most of the contemporary PTA data analyses. 
Adopting a scaling relation between $h_{\rm c}$ and astrophysical parameters $(a_\bullet,b_\bullet,\rho_{\rm BH},M_*,\sigma_*)$ and assuming astrophysically-motivated priors on these parameters, Ref.~\cite{GoncharovSardana2025a} obtains posteriors on these parameters. 
Recasting these parameters in terms of $h_{\rm c}$, which is the quantity fitted to EPTA data, we show $1\sigma$ credible levels of the prior on $h_{\rm c}$ in Figure~\ref{fig:rho_mpeak} as two dashed lines with constant $h_{\rm c}$. 
The posterior matches well with our full posterior on the NANOGrav data, suggesting a consistent astrophysical interpretation between NANOGrav and the EPTA. 
It also illustrates the ability of the model introduced in Ref.~\cite{Sato-PolitoZaldarriaga2025} and our extension of this model to break the degeneracy between $\rho_{\rm BH}$ and $M_{\rm peak}$. 

We also extend the scaling relation used in EPTA data analysis to fully recast prior samples on $(a_\bullet,b_\bullet,\rho_{\rm BH},M_*,\sigma_*)$ from the EPTA analysis~\cite{GoncharovSardana2025a} in terms of $\rho_{\rm BH}$ and $M_{\rm peak}$. 
The priors on $(a_\bullet,b_\bullet)$ are obtained from kinematic observations of local black holes in galaxy catalogues~\cite{McConnellMa2013}. 
The prior on $\sigma_*$ is based on the velocity dispersion of galaxies measured in the Sloan Digital Sky Survey~\cite{BernardiShankar2010}. 
The prior on $\rho_{\rm BH}$ is obtained from Ref.~\cite{LiepoldMa2024}.
The resulting $1\sigma$ level is shown as the thick brown contour in Figure~\ref{fig:rho_mpeak}. 
The choice of Ref.~\cite{LiepoldMa2024} for $\rho_{\rm BH}$ leads to the overlap with the plotted NANOGrav and EPTA observations. 
Adopting the prior on $\rho_{\rm BH}$ from Ref.~\cite{Sato-PolitoZaldarriaga2025} would push the thick brown curve down to the level of $\rho_{\rm BH,fid} = 4.53\times 10^5 M_\odot {\rm Mpc}^{-3}$. 

From the observational perspective, the brown contour in Figure~\ref{fig:rho_mpeak} reveals the following. 
Our new data analysis method adds a new dimension to the inference of SMBHB properties, making the data analysis procedure more complete and correct. 
Furthermore, it shows that the astrophysically expected $N_{\rm c}$ value is within $1\sigma$ of our posterior. 
The brown contour suggests a fairly strong level of non-Gaussianity at about $N_{\rm c}=1$ predicted by this specific choice of prior, providing an optimistic outline for confirming the SMBHB origin of the GWB.
In contrast, $N_{\rm c}=1000$ at the edge of our prior represents a strong level of Gaussianity, which would make SMBHB-related discreteness in the GWB power spectrum much more difficult to resolve. 
If the posterior for future observations narrows down on the thick brown contour, it would confirm that the GWB amplitude and its spectral fluctuations are consistent with the mass scale of local, massive black holes and abundance implied by galaxy demographics.

\section{\label{sec:discussion}Discussion}

We summarize the astrophysical observables specific to the astrophysical GWB from SMBHBs in Table~\ref{tab:hypothesis_hierarchy}. 
The analysis we performed in this work is based on the assumption of circular binaries, GW-driven frequency evolution for SMBHBs above $T_{\rm obs}^{-1}$, and the invariance of $\mu(x)$ to population properties of SMBHBs. 
Effects of black hole spins~\cite{SesanaBarausse2014}, orbital eccentricity~\cite{FalxaLeclere2025}, and SMBHB environment on inspirals can alter the canonical $h_{\rm c}\propto f^{-2/3}$ dependence~\cite{RaviWyithe2014} and introduce other effects such as apparent non-stationarity.
These effects diminish the background's amplitude at low frequencies, so they are not guaranteed to be measurable because the inferred background amplitude appears exceptionally high~\cite{Sato-PolitoZaldarriaga2024}.
Nevertheless, it is important to include the treatment of these effects in our model in the future. 

\begin{table*}
\centering
\small
\caption{Towards an identification of the origin of the GWB with hierarchical models.}
\label{tab:hypothesis_hierarchy}
\begingroup
\renewcommand{\arraystretch}{1.18}
\begin{tabular}{|p{0.22\textwidth}|p{0.27\textwidth}|p{0.22\textwidth}|p{0.21\textwidth}|}
\hline
\textbf{Hypothesis} & \textbf{Observable} & \textbf{Interpretation} & \textbf{Status} \\
\hline
CRN
&
Common power spectrum component in the data of all pulsars
&
Strain power spectrum
&
Verified~\cite{GoncharovThrane2022}
\\
\hline
GWB
&
Hellings--Downs (HD) correlations
&
Tensorial GW polarizations of General Relativity
&
Strong evidence~\cite{NG15_GWB,EPTA_DR2_GW,CPTA_DR1_GWB}
\\
\hline
Astrophysical GWB
&
Finite $N_{\rm c}$
&
SMBHB origin of the GWB
&
Developed here
\\
\hline
\hspace*{1em}\emph{Canonical}
&
$h_{\rm c}(f) \propto f^{-2/3}$
&
Circular orbits, GW-driven SMBHB inspirals
&
Assumed here, consistent with observations~\cite{GoncharovSardana2025a}
\\
\hline
\hspace*{1em}\emph{Non-canonical}
&
$h_{\rm c}(f)$ with a spectral turnover
&
Environmental coupling, eccentricity, or non-canonical SMBHB evolution
&
Requires an extension of the $h_{\rm c}$ scaling assumed in this work
\\
\hline
GWB with a bright source
&
CW
&
An SMBHB signal
&
Developed here
\\
\hline
\hspace*{1em}\emph{Adiabatic SMBHB}
&
Evidence for the Earth term (Pulsar term is optional)
&
SMBHB-associated amplitude and phase modulations
&
~
\\
\hline
\hspace*{1em}\emph{Evolving SMBHB}
&
Evidence for the Pulsar term on top of the Earth term
&
SMBHB-associated amplitude and phase evolution
&
~
\\
\hline
Cosmological GWB
&
Deviations from the above
&
Refs.~\cite{EPTA_DR2_NEWPHYS,NG15_NEWPHYS}
&
Not considered here
\\
\hline
\end{tabular}
\endgroup
\end{table*}


\section{\label{sec:conclusion}Conclusion}

We introduced a hierarchical model for PTA data analysis in which the unresolved GWB and the brightest SMBHB sources are described as two components of the same astrophysical population.
The model extends the SMBHB strain-distribution calculation of Ref.~\cite{Sato-PolitoZaldarriaga2025} to a joint search for the GWB and an individually resolvable CW, allowing the data to simultaneously inform on the inferred background spectrum, CW strain limits, and SMBHB population parameters.
Applied to the simulated NANOGrav 15-year data~\cite{NG15_data,NG15_GWB,NG15_SMBHB}, the analysis finds the data to be insensitive to both individually resolvable SMBHB and the SMBHB-specific Poisson regime of the GWB. 
The absence of a resolvable CW, as per the simulated model, is consistent with the broad posterior support for $f_{\rm cw}$, the agreement between posterior and posterior-predictive samples of $h_{\rm cw}$, and the low posterior-predictive probability of exceeding a fiducial CW detection threshold. 
We propose these criteria for the checklist of CW detection. 
Furthermore, our hierarchical prior on the CW strain amplitude leads to stronger astrophysical limits on the CW strain amplitude than an astrophysically agnostic all-sky search, placing a larger number of AGN-selected SMBHB candidates in tension with the NANOGrav data~\cite{NG15_targeted}. 
This is consistent with the previous findings, suggesting that the considered SMBHB candidates are incompatible with the inferred GWB strain amplitude~\cite{Casey-ClydeMingarelli2025,SesanaHaiman2018}. 
When applied to the real data, our methods could test whether the low-significance spectral and CW outliers are consistent with being genuine resolvable CW signals from SMBHBs.
Recasting the posterior on $(N_{\rm c},h_{\rm c})$ into $(\rho_{\rm BH},M_{\rm peak})$ shows that the PTA results can be compared directly with population information from local SMBH--galaxy scaling relations and with EPTA-based astrophysical inference~\cite{McConnellMa2013,BernardiShankar2010,LiepoldMa2024,GoncharovSardana2025a}.
Future PTA data can therefore test the SMBHB interpretation of the GWB in a more specific way than by measuring its amplitude alone: a narrowing of the posterior toward the astrophysically expected region in $(\rho_{\rm BH},M_{\rm peak})$ and toward smaller $N_{\rm c}$ would indicate the onset of source-count fluctuations expected from a finite SMBHB population.

\section*{Acknowledgements}

This research was supported in part by grant NSF PHY-2309135 to the Kavli Institute for Theoretical Physics (KITP). GSP acknowledges support from the Friends of the Institute for Advanced Study Fund. MZ acknowledges support from the National Science Foundation NSF-BSF 2207583 and NSF 2209991 and the Nelson Center for Collaborative Research and the Simons Foundation through Award No. SFI-MPS-BH-00012593-10.

\bibliography{mybib,collab_papers,soft}{}

\begin{thebibliography}{52}%
\makeatletter
\providecommand \@ifxundefined [1]{%
 \@ifx{#1\undefined}
}%
\providecommand \@ifnum [1]{%
 \ifnum #1\expandafter \@firstoftwo
 \else \expandafter \@secondoftwo
 \fi
}%
\providecommand \@ifx [1]{%
 \ifx #1\expandafter \@firstoftwo
 \else \expandafter \@secondoftwo
 \fi
}%
\providecommand \natexlab [1]{#1}%
\providecommand \enquote  [1]{``#1''}%
\providecommand \bibnamefont  [1]{#1}%
\providecommand \bibfnamefont [1]{#1}%
\providecommand \citenamefont [1]{#1}%
\providecommand \href@noop [0]{\@secondoftwo}%
\providecommand \href [0]{\begingroup \@sanitize@url \@href}%
\providecommand \@href[1]{\@@startlink{#1}\@@href}%
\providecommand \@@href[1]{\endgroup#1\@@endlink}%
\providecommand \@sanitize@url [0]{\catcode `\\12\catcode `\$12\catcode
  `\&12\catcode `\#12\catcode `\^12\catcode `\_12\catcode `\%12\relax}%
\providecommand \@@startlink[1]{}%
\providecommand \@@endlink[0]{}%
\providecommand \url  [0]{\begingroup\@sanitize@url \@url }%
\providecommand \@url [1]{\endgroup\@href {#1}{\urlprefix }}%
\providecommand \urlprefix  [0]{URL }%
\providecommand \Eprint [0]{\href }%
\providecommand \doibase [0]{https://doi.org/}%
\providecommand \selectlanguage [0]{\@gobble}%
\providecommand \bibinfo  [0]{\@secondoftwo}%
\providecommand \bibfield  [0]{\@secondoftwo}%
\providecommand \translation [1]{[#1]}%
\providecommand \BibitemOpen [0]{}%
\providecommand \bibitemStop [0]{}%
\providecommand \bibitemNoStop [0]{.\EOS\space}%
\providecommand \EOS [0]{\spacefactor3000\relax}%
\providecommand \BibitemShut  [1]{\csname bibitem#1\endcsname}%
\let\auto@bib@innerbib\@empty
\bibitem [{\citenamefont {{Nanograv
  Collaboration}}(2023{\natexlab{a}})}]{NG15_GWB}%
  \BibitemOpen
  \bibfield  {author} {\bibinfo {author} {\bibnamefont {{Nanograv
  Collaboration}}},\ }\href {https://doi.org/10.3847/2041-8213/acdac6}
  {\bibfield  {journal} {\bibinfo  {journal} {\apjl}\ }\textbf {\bibinfo
  {volume} {951}},\ \bibinfo {eid} {L8} (\bibinfo {year}
  {2023}{\natexlab{a}})},\ \Eprint {https://arxiv.org/abs/2306.16213}
  {arXiv:2306.16213 [astro-ph.HE]} \BibitemShut {NoStop}%
\bibitem [{\citenamefont {{EPTA Collaboration}}\ and\ \citenamefont {{InPTA
  Collaboration}}(2023)}]{EPTA_DR2_GW}%
  \BibitemOpen
  \bibfield  {author} {\bibinfo {author} {\bibnamefont {{EPTA Collaboration}}}\
  and\ \bibinfo {author} {\bibnamefont {{InPTA Collaboration}}},\ }\href
  {https://doi.org/10.1051/0004-6361/202346844} {\bibfield  {journal} {\bibinfo
   {journal} {\aap}\ }\textbf {\bibinfo {volume} {678}},\ \bibinfo {eid} {A50}
  (\bibinfo {year} {2023})},\ \Eprint {https://arxiv.org/abs/2306.16214}
  {arXiv:2306.16214 [astro-ph.HE]} \BibitemShut {NoStop}%
\bibitem [{\citenamefont {{Reardon}}\ \emph {et~al.}(2023)\citenamefont
  {{Reardon}}, \citenamefont {{Zic}}, \citenamefont {{Shannon}}, \citenamefont
  {{Hobbs}}, \citenamefont {{Bailes}}, \citenamefont {{Di Marco}},
  \citenamefont {{Kapur}}, \citenamefont {{Rogers}}, \citenamefont {{Thrane}},
  \citenamefont {{Askew}}, \citenamefont {{Bhat}}, \citenamefont {{Cameron}},
  \citenamefont {{Cury{\l}o}}, \citenamefont {{Coles}}, \citenamefont {{Dai}},
  \citenamefont {{Goncharov}}, \citenamefont {{Kerr}}, \citenamefont
  {{Kulkarni}}, \citenamefont {{Levin}}, \citenamefont {{Lower}}, \citenamefont
  {{Manchester}}, \citenamefont {{Mandow}}, \citenamefont {{Miles}},
  \citenamefont {{Nathan}}, \citenamefont {{Os{\l}owski}}, \citenamefont
  {{Russell}}, \citenamefont {{Spiewak}}, \citenamefont {{Zhang}},\ and\
  \citenamefont {{Zhu}}}]{PPTA_DR3_GWB}%
  \BibitemOpen
  \bibfield  {author} {\bibinfo {author} {\bibfnamefont {D.~J.}\ \bibnamefont
  {{Reardon}}}, \bibinfo {author} {\bibfnamefont {A.}~\bibnamefont {{Zic}}},
  \bibinfo {author} {\bibfnamefont {R.~M.}\ \bibnamefont {{Shannon}}}, \bibinfo
  {author} {\bibfnamefont {G.~B.}\ \bibnamefont {{Hobbs}}}, \bibinfo {author}
  {\bibfnamefont {M.}~\bibnamefont {{Bailes}}}, \bibinfo {author}
  {\bibfnamefont {V.}~\bibnamefont {{Di Marco}}}, \bibinfo {author}
  {\bibfnamefont {A.}~\bibnamefont {{Kapur}}}, \bibinfo {author} {\bibfnamefont
  {A.~F.}\ \bibnamefont {{Rogers}}}, \bibinfo {author} {\bibfnamefont
  {E.}~\bibnamefont {{Thrane}}}, \bibinfo {author} {\bibfnamefont
  {J.}~\bibnamefont {{Askew}}}, \bibinfo {author} {\bibfnamefont {N.~D.~R.}\
  \bibnamefont {{Bhat}}}, \bibinfo {author} {\bibfnamefont {A.}~\bibnamefont
  {{Cameron}}}, \bibinfo {author} {\bibfnamefont {M.}~\bibnamefont
  {{Cury{\l}o}}}, \bibinfo {author} {\bibfnamefont {W.~A.}\ \bibnamefont
  {{Coles}}}, \bibinfo {author} {\bibfnamefont {S.}~\bibnamefont {{Dai}}},
  \bibinfo {author} {\bibfnamefont {B.}~\bibnamefont {{Goncharov}}}, \bibinfo
  {author} {\bibfnamefont {M.}~\bibnamefont {{Kerr}}}, \bibinfo {author}
  {\bibfnamefont {A.}~\bibnamefont {{Kulkarni}}}, \bibinfo {author}
  {\bibfnamefont {Y.}~\bibnamefont {{Levin}}}, \bibinfo {author} {\bibfnamefont
  {M.~E.}\ \bibnamefont {{Lower}}}, \bibinfo {author} {\bibfnamefont {R.~N.}\
  \bibnamefont {{Manchester}}}, \bibinfo {author} {\bibfnamefont
  {R.}~\bibnamefont {{Mandow}}}, \bibinfo {author} {\bibfnamefont {M.~T.}\
  \bibnamefont {{Miles}}}, \bibinfo {author} {\bibfnamefont {R.~S.}\
  \bibnamefont {{Nathan}}}, \bibinfo {author} {\bibfnamefont {S.}~\bibnamefont
  {{Os{\l}owski}}}, \bibinfo {author} {\bibfnamefont {C.~J.}\ \bibnamefont
  {{Russell}}}, \bibinfo {author} {\bibfnamefont {R.}~\bibnamefont
  {{Spiewak}}}, \bibinfo {author} {\bibfnamefont {S.}~\bibnamefont {{Zhang}}},\
  and\ \bibinfo {author} {\bibfnamefont {X.-J.}\ \bibnamefont {{Zhu}}},\ }\href
  {https://doi.org/10.3847/2041-8213/acdd02} {\bibfield  {journal} {\bibinfo
  {journal} {\apjl}\ }\textbf {\bibinfo {volume} {951}},\ \bibinfo {eid} {L6}
  (\bibinfo {year} {2023})},\ \Eprint {https://arxiv.org/abs/2306.16215}
  {arXiv:2306.16215 [astro-ph.HE]} \BibitemShut {NoStop}%
\bibitem [{\citenamefont {{Xu}}\ \emph {et~al.}(2023)\citenamefont {{Xu}},
  \citenamefont {{Chen}}, \citenamefont {{Guo}}, \citenamefont {{Jiang}},
  \citenamefont {{Wang}}, \citenamefont {{Xu}}, \citenamefont {{Xue}},
  \citenamefont {{Caballero}}, \citenamefont {{Yuan}}, \citenamefont {{Xu}},
  \citenamefont {{Wang}}, \citenamefont {{Hao}}, \citenamefont {{Luo}},
  \citenamefont {{Lee}}, \citenamefont {{Han}}, \citenamefont {{Jiang}},
  \citenamefont {{Shen}}, \citenamefont {{Wang}}, \citenamefont {{Wang}},
  \citenamefont {{Xu}}, \citenamefont {{Wu}}, \citenamefont {{Manchester}},
  \citenamefont {{Qian}}, \citenamefont {{Guan}}, \citenamefont {{Huang}},
  \citenamefont {{Sun}},\ and\ \citenamefont {{Zhu}}}]{CPTA_DR1_GWB}%
  \BibitemOpen
  \bibfield  {author} {\bibinfo {author} {\bibfnamefont {H.}~\bibnamefont
  {{Xu}}}, \bibinfo {author} {\bibfnamefont {S.}~\bibnamefont {{Chen}}},
  \bibinfo {author} {\bibfnamefont {Y.}~\bibnamefont {{Guo}}}, \bibinfo
  {author} {\bibfnamefont {J.}~\bibnamefont {{Jiang}}}, \bibinfo {author}
  {\bibfnamefont {B.}~\bibnamefont {{Wang}}}, \bibinfo {author} {\bibfnamefont
  {J.}~\bibnamefont {{Xu}}}, \bibinfo {author} {\bibfnamefont {Z.}~\bibnamefont
  {{Xue}}}, \bibinfo {author} {\bibfnamefont {R.~N.}\ \bibnamefont
  {{Caballero}}}, \bibinfo {author} {\bibfnamefont {J.}~\bibnamefont {{Yuan}}},
  \bibinfo {author} {\bibfnamefont {Y.}~\bibnamefont {{Xu}}}, \bibinfo {author}
  {\bibfnamefont {J.}~\bibnamefont {{Wang}}}, \bibinfo {author} {\bibfnamefont
  {L.}~\bibnamefont {{Hao}}}, \bibinfo {author} {\bibfnamefont
  {J.}~\bibnamefont {{Luo}}}, \bibinfo {author} {\bibfnamefont
  {K.}~\bibnamefont {{Lee}}}, \bibinfo {author} {\bibfnamefont
  {J.}~\bibnamefont {{Han}}}, \bibinfo {author} {\bibfnamefont
  {P.}~\bibnamefont {{Jiang}}}, \bibinfo {author} {\bibfnamefont
  {Z.}~\bibnamefont {{Shen}}}, \bibinfo {author} {\bibfnamefont
  {M.}~\bibnamefont {{Wang}}}, \bibinfo {author} {\bibfnamefont
  {N.}~\bibnamefont {{Wang}}}, \bibinfo {author} {\bibfnamefont
  {R.}~\bibnamefont {{Xu}}}, \bibinfo {author} {\bibfnamefont {X.}~\bibnamefont
  {{Wu}}}, \bibinfo {author} {\bibfnamefont {R.}~\bibnamefont {{Manchester}}},
  \bibinfo {author} {\bibfnamefont {L.}~\bibnamefont {{Qian}}}, \bibinfo
  {author} {\bibfnamefont {X.}~\bibnamefont {{Guan}}}, \bibinfo {author}
  {\bibfnamefont {M.}~\bibnamefont {{Huang}}}, \bibinfo {author} {\bibfnamefont
  {C.}~\bibnamefont {{Sun}}},\ and\ \bibinfo {author} {\bibfnamefont
  {Y.}~\bibnamefont {{Zhu}}},\ }\href
  {https://doi.org/10.1088/1674-4527/acdfa5} {\bibfield  {journal} {\bibinfo
  {journal} {Research in Astronomy and Astrophysics}\ }\textbf {\bibinfo
  {volume} {23}},\ \bibinfo {eid} {075024} (\bibinfo {year} {2023})},\ \Eprint
  {https://arxiv.org/abs/2306.16216} {arXiv:2306.16216 [astro-ph.HE]}
  \BibitemShut {NoStop}%
\bibitem [{\citenamefont {{Miles}}\ \emph {et~al.}(2025)\citenamefont
  {{Miles}}, \citenamefont {{Shannon}}, \citenamefont {{Reardon}},
  \citenamefont {{Bailes}}, \citenamefont {{Champion}}, \citenamefont
  {{Geyer}}, \citenamefont {{Gitika}}, \citenamefont {{Grunthal}},
  \citenamefont {{Keith}}, \citenamefont {{Kramer}}, \citenamefont
  {{Kulkarni}}, \citenamefont {{Nathan}}, \citenamefont {{Parthasarathy}},
  \citenamefont {{Singha}}, \citenamefont {{Theureau}}, \citenamefont
  {{Thrane}}, \citenamefont {{Abbate}}, \citenamefont {{Buchner}},
  \citenamefont {{Cameron}}, \citenamefont {{Camilo}}, \citenamefont
  {{Moreschi}}, \citenamefont {{Shaifullah}}, \citenamefont {{Shamohammadi}},
  \citenamefont {{Possenti}},\ and\ \citenamefont {{Krishnan}}}]{MT_DR1_GW}%
  \BibitemOpen
  \bibfield  {author} {\bibinfo {author} {\bibfnamefont {M.~T.}\ \bibnamefont
  {{Miles}}}, \bibinfo {author} {\bibfnamefont {R.~M.}\ \bibnamefont
  {{Shannon}}}, \bibinfo {author} {\bibfnamefont {D.~J.}\ \bibnamefont
  {{Reardon}}}, \bibinfo {author} {\bibfnamefont {M.}~\bibnamefont {{Bailes}}},
  \bibinfo {author} {\bibfnamefont {D.~J.}\ \bibnamefont {{Champion}}},
  \bibinfo {author} {\bibfnamefont {M.}~\bibnamefont {{Geyer}}}, \bibinfo
  {author} {\bibfnamefont {P.}~\bibnamefont {{Gitika}}}, \bibinfo {author}
  {\bibfnamefont {K.}~\bibnamefont {{Grunthal}}}, \bibinfo {author}
  {\bibfnamefont {M.~J.}\ \bibnamefont {{Keith}}}, \bibinfo {author}
  {\bibfnamefont {M.}~\bibnamefont {{Kramer}}}, \bibinfo {author}
  {\bibfnamefont {A.~D.}\ \bibnamefont {{Kulkarni}}}, \bibinfo {author}
  {\bibfnamefont {R.~S.}\ \bibnamefont {{Nathan}}}, \bibinfo {author}
  {\bibfnamefont {A.}~\bibnamefont {{Parthasarathy}}}, \bibinfo {author}
  {\bibfnamefont {J.}~\bibnamefont {{Singha}}}, \bibinfo {author}
  {\bibfnamefont {G.}~\bibnamefont {{Theureau}}}, \bibinfo {author}
  {\bibfnamefont {E.}~\bibnamefont {{Thrane}}}, \bibinfo {author}
  {\bibfnamefont {F.}~\bibnamefont {{Abbate}}}, \bibinfo {author}
  {\bibfnamefont {S.}~\bibnamefont {{Buchner}}}, \bibinfo {author}
  {\bibfnamefont {A.~D.}\ \bibnamefont {{Cameron}}}, \bibinfo {author}
  {\bibfnamefont {F.}~\bibnamefont {{Camilo}}}, \bibinfo {author}
  {\bibfnamefont {B.~E.}\ \bibnamefont {{Moreschi}}}, \bibinfo {author}
  {\bibfnamefont {G.}~\bibnamefont {{Shaifullah}}}, \bibinfo {author}
  {\bibfnamefont {M.}~\bibnamefont {{Shamohammadi}}}, \bibinfo {author}
  {\bibfnamefont {A.}~\bibnamefont {{Possenti}}},\ and\ \bibinfo {author}
  {\bibfnamefont {V.~V.}\ \bibnamefont {{Krishnan}}},\ }\href
  {https://doi.org/10.1093/mnras/stae2571} {\bibfield  {journal} {\bibinfo
  {journal} {\mnras}\ }\textbf {\bibinfo {volume} {536}},\ \bibinfo {pages}
  {1489} (\bibinfo {year} {2025})},\ \Eprint {https://arxiv.org/abs/2412.01153}
  {arXiv:2412.01153 [astro-ph.HE]} \BibitemShut {NoStop}%
\bibitem [{\citenamefont {{Nanograv Collaboration}}(2020)}]{NG12_GWB}%
  \BibitemOpen
  \bibfield  {author} {\bibinfo {author} {\bibnamefont {{Nanograv
  Collaboration}}},\ }\href {https://doi.org/10.3847/2041-8213/abd401}
  {\bibfield  {journal} {\bibinfo  {journal} {\apjl}\ }\textbf {\bibinfo
  {volume} {905}},\ \bibinfo {eid} {L34} (\bibinfo {year} {2020})},\ \Eprint
  {https://arxiv.org/abs/2009.04496} {arXiv:2009.04496 [astro-ph.HE]}
  \BibitemShut {NoStop}%
\bibitem [{\citenamefont {{Goncharov}}\ \emph {et~al.}(2021)\citenamefont
  {{Goncharov}}, \citenamefont {{Shannon}}, \citenamefont {{Reardon}},
  \citenamefont {{Hobbs}}, \citenamefont {{Zic}}, \citenamefont {{Bailes}},
  \citenamefont {{Cury{\l}o}}, \citenamefont {{Dai}}, \citenamefont {{Kerr}},
  \citenamefont {{Lower}}, \citenamefont {{Manchester}}, \citenamefont
  {{Mandow}}, \citenamefont {{Middleton}}, \citenamefont {{Miles}},
  \citenamefont {{Parthasarathy}}, \citenamefont {{Thrane}}, \citenamefont
  {{Thyagarajan}}, \citenamefont {{Xue}}, \citenamefont {{Zhu}}, \citenamefont
  {{Cameron}}, \citenamefont {{Feng}}, \citenamefont {{Luo}}, \citenamefont
  {{Russell}}, \citenamefont {{Sarkissian}}, \citenamefont {{Spiewak}},
  \citenamefont {{Wang}}, \citenamefont {{Wang}}, \citenamefont {{Zhang}},\
  and\ \citenamefont {{Zhang}}}]{GoncharovShannon2021}%
  \BibitemOpen
  \bibfield  {author} {\bibinfo {author} {\bibfnamefont {B.}~\bibnamefont
  {{Goncharov}}}, \bibinfo {author} {\bibfnamefont {R.~M.}\ \bibnamefont
  {{Shannon}}}, \bibinfo {author} {\bibfnamefont {D.~J.}\ \bibnamefont
  {{Reardon}}}, \bibinfo {author} {\bibfnamefont {G.}~\bibnamefont {{Hobbs}}},
  \bibinfo {author} {\bibfnamefont {A.}~\bibnamefont {{Zic}}}, \bibinfo
  {author} {\bibfnamefont {M.}~\bibnamefont {{Bailes}}}, \bibinfo {author}
  {\bibfnamefont {M.}~\bibnamefont {{Cury{\l}o}}}, \bibinfo {author}
  {\bibfnamefont {S.}~\bibnamefont {{Dai}}}, \bibinfo {author} {\bibfnamefont
  {M.}~\bibnamefont {{Kerr}}}, \bibinfo {author} {\bibfnamefont {M.~E.}\
  \bibnamefont {{Lower}}}, \bibinfo {author} {\bibfnamefont {R.~N.}\
  \bibnamefont {{Manchester}}}, \bibinfo {author} {\bibfnamefont
  {R.}~\bibnamefont {{Mandow}}}, \bibinfo {author} {\bibfnamefont
  {H.}~\bibnamefont {{Middleton}}}, \bibinfo {author} {\bibfnamefont {M.~T.}\
  \bibnamefont {{Miles}}}, \bibinfo {author} {\bibfnamefont {A.}~\bibnamefont
  {{Parthasarathy}}}, \bibinfo {author} {\bibfnamefont {E.}~\bibnamefont
  {{Thrane}}}, \bibinfo {author} {\bibfnamefont {N.}~\bibnamefont
  {{Thyagarajan}}}, \bibinfo {author} {\bibfnamefont {X.}~\bibnamefont
  {{Xue}}}, \bibinfo {author} {\bibfnamefont {X.~J.}\ \bibnamefont {{Zhu}}},
  \bibinfo {author} {\bibfnamefont {A.~D.}\ \bibnamefont {{Cameron}}}, \bibinfo
  {author} {\bibfnamefont {Y.}~\bibnamefont {{Feng}}}, \bibinfo {author}
  {\bibfnamefont {R.}~\bibnamefont {{Luo}}}, \bibinfo {author} {\bibfnamefont
  {C.~J.}\ \bibnamefont {{Russell}}}, \bibinfo {author} {\bibfnamefont
  {J.}~\bibnamefont {{Sarkissian}}}, \bibinfo {author} {\bibfnamefont
  {R.}~\bibnamefont {{Spiewak}}}, \bibinfo {author} {\bibfnamefont
  {S.}~\bibnamefont {{Wang}}}, \bibinfo {author} {\bibfnamefont {J.~B.}\
  \bibnamefont {{Wang}}}, \bibinfo {author} {\bibfnamefont {L.}~\bibnamefont
  {{Zhang}}},\ and\ \bibinfo {author} {\bibfnamefont {S.}~\bibnamefont
  {{Zhang}}},\ }\href {https://doi.org/10.3847/2041-8213/ac17f4} {\bibfield
  {journal} {\bibinfo  {journal} {\apjl}\ }\textbf {\bibinfo {volume} {917}},\
  \bibinfo {eid} {L19} (\bibinfo {year} {2021})},\ \Eprint
  {https://arxiv.org/abs/2107.12112} {arXiv:2107.12112 [astro-ph.HE]}
  \BibitemShut {NoStop}%
\bibitem [{\citenamefont {{Chen}}\ \emph {et~al.}(2021)\citenamefont {{Chen}},
  \citenamefont {{Caballero}}, \citenamefont {{Guo}}, \citenamefont
  {{Chalumeau}}, \citenamefont {{Liu}}, \citenamefont {{Shaifullah}},
  \citenamefont {{Lee}}, \citenamefont {{Babak}}, \citenamefont {{Desvignes}},
  \citenamefont {{Parthasarathy}}, \citenamefont {{Hu}}, \citenamefont {{van
  der Wateren}}, \citenamefont {{Antoniadis}}, \citenamefont {{Bak Nielsen}},
  \citenamefont {{Bassa}}, \citenamefont {{Berthereau}}, \citenamefont
  {{Burgay}}, \citenamefont {{Champion}}, \citenamefont {{Cognard}},
  \citenamefont {{Falxa}}, \citenamefont {{Ferdman}}, \citenamefont {{Freire}},
  \citenamefont {{Gair}}, \citenamefont {{Graikou}}, \citenamefont
  {{Guillemot}}, \citenamefont {{Jang}}, \citenamefont {{Janssen}},
  \citenamefont {{Karuppusamy}}, \citenamefont {{Keith}}, \citenamefont
  {{Kramer}}, \citenamefont {{Liu}}, \citenamefont {{Lyne}}, \citenamefont
  {{Main}}, \citenamefont {{McKee}}, \citenamefont {{Mickaliger}},
  \citenamefont {{Perera}}, \citenamefont {{Perrodin}}, \citenamefont
  {{Petiteau}}, \citenamefont {{Porayko}}, \citenamefont {{Possenti}},
  \citenamefont {{Samajdar}}, \citenamefont {{Sanidas}}, \citenamefont
  {{Sesana}}, \citenamefont {{Speri}}, \citenamefont {{Stappers}},
  \citenamefont {{Theureau}}, \citenamefont {{Tiburzi}}, \citenamefont
  {{Vecchio}}, \citenamefont {{Verbiest}}, \citenamefont {{Wang}},
  \citenamefont {{Wang}},\ and\ \citenamefont {{Xu}}}]{ChenCaballero2021}%
  \BibitemOpen
  \bibfield  {author} {\bibinfo {author} {\bibfnamefont {S.}~\bibnamefont
  {{Chen}}}, \bibinfo {author} {\bibfnamefont {R.~N.}\ \bibnamefont
  {{Caballero}}}, \bibinfo {author} {\bibfnamefont {Y.~J.}\ \bibnamefont
  {{Guo}}}, \bibinfo {author} {\bibfnamefont {A.}~\bibnamefont {{Chalumeau}}},
  \bibinfo {author} {\bibfnamefont {K.}~\bibnamefont {{Liu}}}, \bibinfo
  {author} {\bibfnamefont {G.}~\bibnamefont {{Shaifullah}}}, \bibinfo {author}
  {\bibfnamefont {K.~J.}\ \bibnamefont {{Lee}}}, \bibinfo {author}
  {\bibfnamefont {S.}~\bibnamefont {{Babak}}}, \bibinfo {author} {\bibfnamefont
  {G.}~\bibnamefont {{Desvignes}}}, \bibinfo {author} {\bibfnamefont
  {A.}~\bibnamefont {{Parthasarathy}}}, \bibinfo {author} {\bibfnamefont
  {H.}~\bibnamefont {{Hu}}}, \bibinfo {author} {\bibfnamefont {E.}~\bibnamefont
  {{van der Wateren}}}, \bibinfo {author} {\bibfnamefont {J.}~\bibnamefont
  {{Antoniadis}}}, \bibinfo {author} {\bibfnamefont {A.~S.}\ \bibnamefont {{Bak
  Nielsen}}}, \bibinfo {author} {\bibfnamefont {C.~G.}\ \bibnamefont
  {{Bassa}}}, \bibinfo {author} {\bibfnamefont {A.}~\bibnamefont
  {{Berthereau}}}, \bibinfo {author} {\bibfnamefont {M.}~\bibnamefont
  {{Burgay}}}, \bibinfo {author} {\bibfnamefont {D.~J.}\ \bibnamefont
  {{Champion}}}, \bibinfo {author} {\bibfnamefont {I.}~\bibnamefont
  {{Cognard}}}, \bibinfo {author} {\bibfnamefont {M.}~\bibnamefont {{Falxa}}},
  \bibinfo {author} {\bibfnamefont {R.~D.}\ \bibnamefont {{Ferdman}}}, \bibinfo
  {author} {\bibfnamefont {P.~C.~C.}\ \bibnamefont {{Freire}}}, \bibinfo
  {author} {\bibfnamefont {J.~R.}\ \bibnamefont {{Gair}}}, \bibinfo {author}
  {\bibfnamefont {E.}~\bibnamefont {{Graikou}}}, \bibinfo {author}
  {\bibfnamefont {L.}~\bibnamefont {{Guillemot}}}, \bibinfo {author}
  {\bibfnamefont {J.}~\bibnamefont {{Jang}}}, \bibinfo {author} {\bibfnamefont
  {G.~H.}\ \bibnamefont {{Janssen}}}, \bibinfo {author} {\bibfnamefont
  {R.}~\bibnamefont {{Karuppusamy}}}, \bibinfo {author} {\bibfnamefont {M.~J.}\
  \bibnamefont {{Keith}}}, \bibinfo {author} {\bibfnamefont {M.}~\bibnamefont
  {{Kramer}}}, \bibinfo {author} {\bibfnamefont {X.~J.}\ \bibnamefont {{Liu}}},
  \bibinfo {author} {\bibfnamefont {A.~G.}\ \bibnamefont {{Lyne}}}, \bibinfo
  {author} {\bibfnamefont {R.~A.}\ \bibnamefont {{Main}}}, \bibinfo {author}
  {\bibfnamefont {J.~W.}\ \bibnamefont {{McKee}}}, \bibinfo {author}
  {\bibfnamefont {M.~B.}\ \bibnamefont {{Mickaliger}}}, \bibinfo {author}
  {\bibfnamefont {B.~B.~P.}\ \bibnamefont {{Perera}}}, \bibinfo {author}
  {\bibfnamefont {D.}~\bibnamefont {{Perrodin}}}, \bibinfo {author}
  {\bibfnamefont {A.}~\bibnamefont {{Petiteau}}}, \bibinfo {author}
  {\bibfnamefont {N.~K.}\ \bibnamefont {{Porayko}}}, \bibinfo {author}
  {\bibfnamefont {A.}~\bibnamefont {{Possenti}}}, \bibinfo {author}
  {\bibfnamefont {A.}~\bibnamefont {{Samajdar}}}, \bibinfo {author}
  {\bibfnamefont {S.~A.}\ \bibnamefont {{Sanidas}}}, \bibinfo {author}
  {\bibfnamefont {A.}~\bibnamefont {{Sesana}}}, \bibinfo {author}
  {\bibfnamefont {L.}~\bibnamefont {{Speri}}}, \bibinfo {author} {\bibfnamefont
  {B.~W.}\ \bibnamefont {{Stappers}}}, \bibinfo {author} {\bibfnamefont
  {G.}~\bibnamefont {{Theureau}}}, \bibinfo {author} {\bibfnamefont
  {C.}~\bibnamefont {{Tiburzi}}}, \bibinfo {author} {\bibfnamefont
  {A.}~\bibnamefont {{Vecchio}}}, \bibinfo {author} {\bibfnamefont {J.~P.~W.}\
  \bibnamefont {{Verbiest}}}, \bibinfo {author} {\bibfnamefont
  {J.}~\bibnamefont {{Wang}}}, \bibinfo {author} {\bibfnamefont
  {L.}~\bibnamefont {{Wang}}},\ and\ \bibinfo {author} {\bibfnamefont
  {H.}~\bibnamefont {{Xu}}},\ }\href {https://doi.org/10.1093/mnras/stab2833}
  {\bibfield  {journal} {\bibinfo  {journal} {\mnras}\ }\textbf {\bibinfo
  {volume} {508}},\ \bibinfo {pages} {4970} (\bibinfo {year} {2021})},\ \Eprint
  {https://arxiv.org/abs/2110.13184} {arXiv:2110.13184 [astro-ph.HE]}
  \BibitemShut {NoStop}%
\bibitem [{\citenamefont {{The International Pulsar Timing Array
  Collaboration}}(2022)}]{IPTA_DR2_GWB}%
  \BibitemOpen
  \bibfield  {author} {\bibinfo {author} {\bibnamefont {{The International
  Pulsar Timing Array Collaboration}}},\ }\href
  {https://doi.org/10.1093/mnras/stab3418} {\bibfield  {journal} {\bibinfo
  {journal} {\mnras}\ }\textbf {\bibinfo {volume} {510}},\ \bibinfo {pages}
  {4873} (\bibinfo {year} {2022})},\ \Eprint {https://arxiv.org/abs/2201.03980}
  {arXiv:2201.03980 [astro-ph.HE]} \BibitemShut {NoStop}%
\bibitem [{\citenamefont {{Goncharov}}\ \emph {et~al.}(2022)\citenamefont
  {{Goncharov}}, \citenamefont {{Thrane}}, \citenamefont {{Shannon}},
  \citenamefont {{Harms}}, \citenamefont {{Bhat}}, \citenamefont {{Hobbs}},
  \citenamefont {{Kerr}}, \citenamefont {{Manchester}}, \citenamefont
  {{Reardon}}, \citenamefont {{Russell}}, \citenamefont {{Zhu}},\ and\
  \citenamefont {{Zic}}}]{GoncharovThrane2022}%
  \BibitemOpen
  \bibfield  {author} {\bibinfo {author} {\bibfnamefont {B.}~\bibnamefont
  {{Goncharov}}}, \bibinfo {author} {\bibfnamefont {E.}~\bibnamefont
  {{Thrane}}}, \bibinfo {author} {\bibfnamefont {R.~M.}\ \bibnamefont
  {{Shannon}}}, \bibinfo {author} {\bibfnamefont {J.}~\bibnamefont {{Harms}}},
  \bibinfo {author} {\bibfnamefont {N.~D.~R.}\ \bibnamefont {{Bhat}}}, \bibinfo
  {author} {\bibfnamefont {G.}~\bibnamefont {{Hobbs}}}, \bibinfo {author}
  {\bibfnamefont {M.}~\bibnamefont {{Kerr}}}, \bibinfo {author} {\bibfnamefont
  {R.~N.}\ \bibnamefont {{Manchester}}}, \bibinfo {author} {\bibfnamefont
  {D.~J.}\ \bibnamefont {{Reardon}}}, \bibinfo {author} {\bibfnamefont {C.~J.}\
  \bibnamefont {{Russell}}}, \bibinfo {author} {\bibfnamefont {X.-J.}\
  \bibnamefont {{Zhu}}},\ and\ \bibinfo {author} {\bibfnamefont
  {A.}~\bibnamefont {{Zic}}},\ }\href
  {https://doi.org/10.3847/2041-8213/ac76bb} {\bibfield  {journal} {\bibinfo
  {journal} {\apjl}\ }\textbf {\bibinfo {volume} {932}},\ \bibinfo {eid} {L22}
  (\bibinfo {year} {2022})},\ \Eprint {https://arxiv.org/abs/2206.03766}
  {arXiv:2206.03766 [gr-qc]} \BibitemShut {NoStop}%
\bibitem [{\citenamefont {{Jaffe}}\ and\ \citenamefont
  {{Backer}}(2003)}]{JaffeBacker2003}%
  \BibitemOpen
  \bibfield  {author} {\bibinfo {author} {\bibfnamefont {A.~H.}\ \bibnamefont
  {{Jaffe}}}\ and\ \bibinfo {author} {\bibfnamefont {D.~C.}\ \bibnamefont
  {{Backer}}},\ }\href {https://doi.org/10.1086/345443} {\bibfield  {journal}
  {\bibinfo  {journal} {\apj}\ }\textbf {\bibinfo {volume} {583}},\ \bibinfo
  {pages} {616} (\bibinfo {year} {2003})},\ \Eprint
  {https://arxiv.org/abs/astro-ph/0210148} {arXiv:astro-ph/0210148 [astro-ph]}
  \BibitemShut {NoStop}%
\bibitem [{\citenamefont {{Sesana}}\ and\ \citenamefont
  {{Vecchio}}(2010)}]{SesanaVecchio2010}%
  \BibitemOpen
  \bibfield  {author} {\bibinfo {author} {\bibfnamefont {A.}~\bibnamefont
  {{Sesana}}}\ and\ \bibinfo {author} {\bibfnamefont {A.}~\bibnamefont
  {{Vecchio}}},\ }\href {https://doi.org/10.1088/0264-9381/27/8/084016}
  {\bibfield  {journal} {\bibinfo  {journal} {Classical and Quantum Gravity}\
  }\textbf {\bibinfo {volume} {27}},\ \bibinfo {eid} {084016} (\bibinfo {year}
  {2010})},\ \Eprint {https://arxiv.org/abs/1001.3161} {arXiv:1001.3161
  [astro-ph.CO]} \BibitemShut {NoStop}%
\bibitem [{\citenamefont {{Mingarelli}}\ \emph {et~al.}(2013)\citenamefont
  {{Mingarelli}}, \citenamefont {{Sidery}}, \citenamefont {{Mandel}},\ and\
  \citenamefont {{Vecchio}}}]{MingarelliSidery2013}%
  \BibitemOpen
  \bibfield  {author} {\bibinfo {author} {\bibfnamefont {C.~M.~F.}\
  \bibnamefont {{Mingarelli}}}, \bibinfo {author} {\bibfnamefont
  {T.}~\bibnamefont {{Sidery}}}, \bibinfo {author} {\bibfnamefont
  {I.}~\bibnamefont {{Mandel}}},\ and\ \bibinfo {author} {\bibfnamefont
  {A.}~\bibnamefont {{Vecchio}}},\ }\href
  {https://doi.org/10.1103/PhysRevD.88.062005} {\bibfield  {journal} {\bibinfo
  {journal} {\prd}\ }\textbf {\bibinfo {volume} {88}},\ \bibinfo {eid} {062005}
  (\bibinfo {year} {2013})},\ \Eprint {https://arxiv.org/abs/1306.5394}
  {arXiv:1306.5394 [astro-ph.HE]} \BibitemShut {NoStop}%
\bibitem [{\citenamefont {{Rosado}}\ \emph {et~al.}(2015)\citenamefont
  {{Rosado}}, \citenamefont {{Sesana}},\ and\ \citenamefont
  {{Gair}}}]{RosadoSesana2015}%
  \BibitemOpen
  \bibfield  {author} {\bibinfo {author} {\bibfnamefont {P.~A.}\ \bibnamefont
  {{Rosado}}}, \bibinfo {author} {\bibfnamefont {A.}~\bibnamefont {{Sesana}}},\
  and\ \bibinfo {author} {\bibfnamefont {J.}~\bibnamefont {{Gair}}},\ }\href
  {https://doi.org/10.1093/mnras/stv1098} {\bibfield  {journal} {\bibinfo
  {journal} {\mnras}\ }\textbf {\bibinfo {volume} {451}},\ \bibinfo {pages}
  {2417} (\bibinfo {year} {2015})},\ \Eprint {https://arxiv.org/abs/1503.04803}
  {arXiv:1503.04803 [astro-ph.HE]} \BibitemShut {NoStop}%
\bibitem [{\citenamefont {{Nanograv
  Collaboration}}(2023{\natexlab{b}})}]{NG15_SMBHB}%
  \BibitemOpen
  \bibfield  {author} {\bibinfo {author} {\bibnamefont {{Nanograv
  Collaboration}}},\ }\href {https://doi.org/10.3847/2041-8213/ace18b}
  {\bibfield  {journal} {\bibinfo  {journal} {\apjl}\ }\textbf {\bibinfo
  {volume} {952}},\ \bibinfo {eid} {L37} (\bibinfo {year}
  {2023}{\natexlab{b}})},\ \Eprint {https://arxiv.org/abs/2306.16220}
  {arXiv:2306.16220 [astro-ph.HE]} \BibitemShut {NoStop}%
\bibitem [{\citenamefont {{Kelley}}\ \emph {et~al.}(2017)\citenamefont
  {{Kelley}}, \citenamefont {{Blecha}}, \citenamefont {{Hernquist}},
  \citenamefont {{Sesana}},\ and\ \citenamefont {{Taylor}}}]{KelleyBlecha2017}%
  \BibitemOpen
  \bibfield  {author} {\bibinfo {author} {\bibfnamefont {L.~Z.}\ \bibnamefont
  {{Kelley}}}, \bibinfo {author} {\bibfnamefont {L.}~\bibnamefont {{Blecha}}},
  \bibinfo {author} {\bibfnamefont {L.}~\bibnamefont {{Hernquist}}}, \bibinfo
  {author} {\bibfnamefont {A.}~\bibnamefont {{Sesana}}},\ and\ \bibinfo
  {author} {\bibfnamefont {S.~R.}\ \bibnamefont {{Taylor}}},\ }\href
  {https://doi.org/10.1093/mnras/stx1638} {\bibfield  {journal} {\bibinfo
  {journal} {\mnras}\ }\textbf {\bibinfo {volume} {471}},\ \bibinfo {pages}
  {4508} (\bibinfo {year} {2017})},\ \Eprint {https://arxiv.org/abs/1702.02180}
  {arXiv:1702.02180 [astro-ph.HE]} \BibitemShut {NoStop}%
\bibitem [{\citenamefont {{Middleton}}\ \emph {et~al.}(2018)\citenamefont
  {{Middleton}}, \citenamefont {{Chen}}, \citenamefont {{Del Pozzo}},
  \citenamefont {{Sesana}},\ and\ \citenamefont
  {{Vecchio}}}]{MiddletonChen2018}%
  \BibitemOpen
  \bibfield  {author} {\bibinfo {author} {\bibfnamefont {H.}~\bibnamefont
  {{Middleton}}}, \bibinfo {author} {\bibfnamefont {S.}~\bibnamefont {{Chen}}},
  \bibinfo {author} {\bibfnamefont {W.}~\bibnamefont {{Del Pozzo}}}, \bibinfo
  {author} {\bibfnamefont {A.}~\bibnamefont {{Sesana}}},\ and\ \bibinfo
  {author} {\bibfnamefont {A.}~\bibnamefont {{Vecchio}}},\ }\href
  {https://doi.org/10.1038/s41467-018-02916-7} {\bibfield  {journal} {\bibinfo
  {journal} {Nature Communications}\ }\textbf {\bibinfo {volume} {9}},\
  \bibinfo {eid} {573} (\bibinfo {year} {2018})},\ \Eprint
  {https://arxiv.org/abs/1707.00623} {arXiv:1707.00623 [astro-ph.GA]}
  \BibitemShut {NoStop}%
\bibitem [{\citenamefont {{Chen}}\ \emph {et~al.}(2019)\citenamefont {{Chen}},
  \citenamefont {{Sesana}},\ and\ \citenamefont
  {{Conselice}}}]{ChenSesana2019}%
  \BibitemOpen
  \bibfield  {author} {\bibinfo {author} {\bibfnamefont {S.}~\bibnamefont
  {{Chen}}}, \bibinfo {author} {\bibfnamefont {A.}~\bibnamefont {{Sesana}}},\
  and\ \bibinfo {author} {\bibfnamefont {C.~J.}\ \bibnamefont {{Conselice}}},\
  }\href {https://doi.org/10.1093/mnras/stz1722} {\bibfield  {journal}
  {\bibinfo  {journal} {\mnras}\ }\textbf {\bibinfo {volume} {488}},\ \bibinfo
  {pages} {401} (\bibinfo {year} {2019})},\ \Eprint
  {https://arxiv.org/abs/1810.04184} {arXiv:1810.04184 [astro-ph.GA]}
  \BibitemShut {NoStop}%
\bibitem [{\citenamefont {{Nanograv
  Collaboration}}(2023{\natexlab{c}})}]{NG15_HOLODECK}%
  \BibitemOpen
  \bibfield  {author} {\bibinfo {author} {\bibnamefont {{Nanograv
  Collaboration}}},\ }\href {https://doi.org/10.3847/2041-8213/ace18b}
  {\bibfield  {journal} {\bibinfo  {journal} {\apjl}\ }\textbf {\bibinfo
  {volume} {952}},\ \bibinfo {eid} {L37} (\bibinfo {year}
  {2023}{\natexlab{c}})},\ \Eprint {https://arxiv.org/abs/2306.16220}
  {arXiv:2306.16220 [astro-ph.HE]} \BibitemShut {NoStop}%
\bibitem [{\citenamefont {{Hellings}}\ and\ \citenamefont
  {{Downs}}(1983)}]{HellingsDowns1983}%
  \BibitemOpen
  \bibfield  {author} {\bibinfo {author} {\bibfnamefont {R.~W.}\ \bibnamefont
  {{Hellings}}}\ and\ \bibinfo {author} {\bibfnamefont {G.~S.}\ \bibnamefont
  {{Downs}}},\ }in\ \href@noop {} {\emph {\bibinfo {booktitle} {General
  Relativity and Gravitation, Volume 1}}},\ Vol.~\bibinfo {volume} {1},\
  \bibinfo {editor} {edited by\ \bibinfo {editor} {\bibfnamefont
  {B.}~\bibnamefont {{Bertotti}}}, \bibinfo {editor} {\bibfnamefont
  {F.}~\bibnamefont {{de Felice}}},\ and\ \bibinfo {editor} {\bibfnamefont
  {A.}~\bibnamefont {{Pascolini}}}}\ (\bibinfo {year} {1983})\ p.\ \bibinfo
  {pages} {963}\BibitemShut {NoStop}%
\bibitem [{\citenamefont {{Sato-Polito}}\ and\ \citenamefont
  {{Zaldarriaga}}(2025)}]{Sato-PolitoZaldarriaga2025}%
  \BibitemOpen
  \bibfield  {author} {\bibinfo {author} {\bibfnamefont {G.}~\bibnamefont
  {{Sato-Polito}}}\ and\ \bibinfo {author} {\bibfnamefont {M.}~\bibnamefont
  {{Zaldarriaga}}},\ }\href {https://doi.org/10.1103/PhysRevD.111.023043}
  {\bibfield  {journal} {\bibinfo  {journal} {\prd}\ }\textbf {\bibinfo
  {volume} {111}},\ \bibinfo {eid} {023043} (\bibinfo {year} {2025})},\ \Eprint
  {https://arxiv.org/abs/2406.17010} {arXiv:2406.17010 [astro-ph.CO]}
  \BibitemShut {NoStop}%
\bibitem [{\citenamefont {{Xue}}\ \emph {et~al.}(2025)\citenamefont {{Xue}},
  \citenamefont {{Pan}},\ and\ \citenamefont {{Dai}}}]{XuePan2025}%
  \BibitemOpen
  \bibfield  {author} {\bibinfo {author} {\bibfnamefont {X.}~\bibnamefont
  {{Xue}}}, \bibinfo {author} {\bibfnamefont {Z.}~\bibnamefont {{Pan}}},\ and\
  \bibinfo {author} {\bibfnamefont {L.}~\bibnamefont {{Dai}}},\ }\href
  {https://doi.org/10.1103/PhysRevD.111.043022} {\bibfield  {journal} {\bibinfo
   {journal} {\prd}\ }\textbf {\bibinfo {volume} {111}},\ \bibinfo {eid}
  {043022} (\bibinfo {year} {2025})},\ \Eprint
  {https://arxiv.org/abs/2409.19516} {arXiv:2409.19516 [astro-ph.CO]}
  \BibitemShut {NoStop}%
\bibitem [{\citenamefont {{Wu}}\ \emph {et~al.}(2024)\citenamefont {{Wu}},
  \citenamefont {{Bi}},\ and\ \citenamefont {{Huang}}}]{WuBi2024}%
  \BibitemOpen
  \bibfield  {author} {\bibinfo {author} {\bibfnamefont {Y.-M.}\ \bibnamefont
  {{Wu}}}, \bibinfo {author} {\bibfnamefont {Y.-C.}\ \bibnamefont {{Bi}}},\
  and\ \bibinfo {author} {\bibfnamefont {Q.-G.}\ \bibnamefont {{Huang}}},\
  }\href {https://doi.org/10.48550/arXiv.2407.07319} {\bibfield  {journal}
  {\bibinfo  {journal} {arXiv e-prints}\ ,\ \bibinfo {eid} {arXiv:2407.07319}}
  (\bibinfo {year} {2024})},\ \Eprint {https://arxiv.org/abs/2407.07319}
  {arXiv:2407.07319 [astro-ph.CO]} \BibitemShut {NoStop}%
\bibitem [{\citenamefont {{Ellis}}\ \emph {et~al.}(2023)\citenamefont
  {{Ellis}}, \citenamefont {{Fairbairn}}, \citenamefont {{H{\"u}tsi}},
  \citenamefont {{Raidal}}, \citenamefont {{Urrutia}}, \citenamefont
  {{Vaskonen}},\ and\ \citenamefont {{Veerm{\"a}e}}}]{EllisFairbairn2023}%
  \BibitemOpen
  \bibfield  {author} {\bibinfo {author} {\bibfnamefont {J.}~\bibnamefont
  {{Ellis}}}, \bibinfo {author} {\bibfnamefont {M.}~\bibnamefont
  {{Fairbairn}}}, \bibinfo {author} {\bibfnamefont {G.}~\bibnamefont
  {{H{\"u}tsi}}}, \bibinfo {author} {\bibfnamefont {M.}~\bibnamefont
  {{Raidal}}}, \bibinfo {author} {\bibfnamefont {J.}~\bibnamefont {{Urrutia}}},
  \bibinfo {author} {\bibfnamefont {V.}~\bibnamefont {{Vaskonen}}},\ and\
  \bibinfo {author} {\bibfnamefont {H.}~\bibnamefont {{Veerm{\"a}e}}},\ }\href
  {https://doi.org/10.1051/0004-6361/202346268} {\bibfield  {journal} {\bibinfo
   {journal} {\aap}\ }\textbf {\bibinfo {volume} {676}},\ \bibinfo {eid} {A38}
  (\bibinfo {year} {2023})},\ \Eprint {https://arxiv.org/abs/2301.13854}
  {arXiv:2301.13854 [astro-ph.CO]} \BibitemShut {NoStop}%
\bibitem [{\citenamefont {{Lamb}}\ and\ \citenamefont
  {{Taylor}}(2024)}]{LambTaylor2024}%
  \BibitemOpen
  \bibfield  {author} {\bibinfo {author} {\bibfnamefont {W.~G.}\ \bibnamefont
  {{Lamb}}}\ and\ \bibinfo {author} {\bibfnamefont {S.~R.}\ \bibnamefont
  {{Taylor}}},\ }\href {https://doi.org/10.3847/2041-8213/ad654a} {\bibfield
  {journal} {\bibinfo  {journal} {\apjl}\ }\textbf {\bibinfo {volume} {971}},\
  \bibinfo {eid} {L10} (\bibinfo {year} {2024})},\ \Eprint
  {https://arxiv.org/abs/2407.06270} {arXiv:2407.06270 [gr-qc]} \BibitemShut
  {NoStop}%
\bibitem [{\citenamefont {{Hisamatsu}}\ and\ \citenamefont
  {{Kyutoku}}(2026)}]{HisamatsuKyutoku2026}%
  \BibitemOpen
  \bibfield  {author} {\bibinfo {author} {\bibfnamefont {H.}~\bibnamefont
  {{Hisamatsu}}}\ and\ \bibinfo {author} {\bibfnamefont {K.}~\bibnamefont
  {{Kyutoku}}},\ }\href {https://doi.org/10.48550/arXiv.2605.17983} {\bibfield
  {journal} {\bibinfo  {journal} {arXiv e-prints}\ ,\ \bibinfo {eid}
  {arXiv:2605.17983}} (\bibinfo {year} {2026})},\ \Eprint
  {https://arxiv.org/abs/2605.17983} {arXiv:2605.17983 [astro-ph.HE]}
  \BibitemShut {NoStop}%
\bibitem [{\citenamefont {{Falxa}}\ and\ \citenamefont
  {{Sesana}}(2026)}]{FalxaSesana2026}%
  \BibitemOpen
  \bibfield  {author} {\bibinfo {author} {\bibfnamefont {M.}~\bibnamefont
  {{Falxa}}}\ and\ \bibinfo {author} {\bibfnamefont {A.}~\bibnamefont
  {{Sesana}}},\ }\href {https://doi.org/10.1103/9jdd-6ct8} {\bibfield
  {journal} {\bibinfo  {journal} {\prd}\ }\textbf {\bibinfo {volume} {113}},\
  \bibinfo {eid} {043047} (\bibinfo {year} {2026})},\ \Eprint
  {https://arxiv.org/abs/2508.08365} {arXiv:2508.08365 [astro-ph.IM]}
  \BibitemShut {NoStop}%
\bibitem [{\citenamefont {{Ali-Ha{\"\i}moud}}(2026)}]{Ali-Haimoud2026}%
  \BibitemOpen
  \bibfield  {author} {\bibinfo {author} {\bibfnamefont {Y.}~\bibnamefont
  {{Ali-Ha{\"\i}moud}}},\ }\href {https://doi.org/10.48550/arXiv.2604.19701}
  {\bibfield  {journal} {\bibinfo  {journal} {arXiv e-prints}\ ,\ \bibinfo
  {eid} {arXiv:2604.19701}} (\bibinfo {year} {2026})},\ \Eprint
  {https://arxiv.org/abs/2604.19701} {arXiv:2604.19701 [astro-ph.CO]}
  \BibitemShut {NoStop}%
\bibitem [{\citenamefont {{Sardesai}}\ \emph {et~al.}(2024)\citenamefont
  {{Sardesai}}, \citenamefont {{Simon}},\ and\ \citenamefont
  {{Vigeland}}}]{SardesaiSimon2024}%
  \BibitemOpen
  \bibfield  {author} {\bibinfo {author} {\bibfnamefont {S.~C.}\ \bibnamefont
  {{Sardesai}}}, \bibinfo {author} {\bibfnamefont {J.}~\bibnamefont
  {{Simon}}},\ and\ \bibinfo {author} {\bibfnamefont {S.~J.}\ \bibnamefont
  {{Vigeland}}},\ }\href {https://doi.org/10.3847/1538-4357/ad8a60} {\bibfield
  {journal} {\bibinfo  {journal} {\apj}\ }\textbf {\bibinfo {volume} {976}},\
  \bibinfo {eid} {212} (\bibinfo {year} {2024})},\ \Eprint
  {https://arxiv.org/abs/2408.10139} {arXiv:2408.10139 [astro-ph.HE]}
  \BibitemShut {NoStop}%
\bibitem [{\citenamefont {{Nanograv
  Collaboration}}(2023{\natexlab{d}})}]{NG15_data}%
  \BibitemOpen
  \bibfield  {author} {\bibinfo {author} {\bibnamefont {{Nanograv
  Collaboration}}},\ }\href {https://doi.org/10.3847/2041-8213/acda9a}
  {\bibfield  {journal} {\bibinfo  {journal} {\apjl}\ }\textbf {\bibinfo
  {volume} {951}},\ \bibinfo {eid} {L9} (\bibinfo {year}
  {2023}{\natexlab{d}})},\ \Eprint {https://arxiv.org/abs/2306.16217}
  {arXiv:2306.16217 [astro-ph.HE]} \BibitemShut {NoStop}%
\bibitem [{\citenamefont {{van Haasteren}}\ \emph {et~al.}(2009)\citenamefont
  {{van Haasteren}}, \citenamefont {{Levin}}, \citenamefont {{McDonald}},\ and\
  \citenamefont {{Lu}}}]{vanHaasterenLevin2009}%
  \BibitemOpen
  \bibfield  {author} {\bibinfo {author} {\bibfnamefont {R.}~\bibnamefont {{van
  Haasteren}}}, \bibinfo {author} {\bibfnamefont {Y.}~\bibnamefont {{Levin}}},
  \bibinfo {author} {\bibfnamefont {P.}~\bibnamefont {{McDonald}}},\ and\
  \bibinfo {author} {\bibfnamefont {T.}~\bibnamefont {{Lu}}},\ }\href
  {https://doi.org/10.1111/j.1365-2966.2009.14590.x} {\bibfield  {journal}
  {\bibinfo  {journal} {\mnras}\ }\textbf {\bibinfo {volume} {395}},\ \bibinfo
  {pages} {1005} (\bibinfo {year} {2009})},\ \Eprint
  {https://arxiv.org/abs/0809.0791} {arXiv:0809.0791 [astro-ph]} \BibitemShut
  {NoStop}%
\bibitem [{\citenamefont {{NANOGrav Collaboration}}(2016)}]{NG9_GWB}%
  \BibitemOpen
  \bibfield  {author} {\bibinfo {author} {\bibnamefont {{NANOGrav
  Collaboration}}},\ }\href {https://doi.org/10.3847/0004-637X/821/1/13}
  {\bibfield  {journal} {\bibinfo  {journal} {\apj}\ }\textbf {\bibinfo
  {volume} {821}},\ \bibinfo {eid} {13} (\bibinfo {year} {2016})},\ \Eprint
  {https://arxiv.org/abs/1508.03024} {arXiv:1508.03024 [astro-ph.GA]}
  \BibitemShut {NoStop}%
\bibitem [{\citenamefont {{NANOGrav Collaboration}}(2023)}]{NG15_NEWPHYS}%
  \BibitemOpen
  \bibfield  {author} {\bibinfo {author} {\bibnamefont {{NANOGrav
  Collaboration}}},\ }\href {https://doi.org/10.3847/2041-8213/acdc91}
  {\bibfield  {journal} {\bibinfo  {journal} {\apjl}\ }\textbf {\bibinfo
  {volume} {951}},\ \bibinfo {eid} {L11} (\bibinfo {year} {2023})},\ \Eprint
  {https://arxiv.org/abs/2306.16219} {arXiv:2306.16219 [astro-ph.HE]}
  \BibitemShut {NoStop}%
\bibitem [{\citenamefont {{Verbiest}}\ \emph {et~al.}(2016)\citenamefont
  {{Verbiest}}, \citenamefont {{Lentati}}, \citenamefont {{Hobbs}},
  \citenamefont {{van Haasteren}}, \citenamefont {{Demorest}}, \citenamefont
  {{Janssen}}, \citenamefont {{Wang}}, \citenamefont {{Desvignes}},
  \citenamefont {{Caballero}}, \citenamefont {{Keith}}, \citenamefont
  {{Champion}} \emph {et~al.}}]{IPTA_DR1}%
  \BibitemOpen
  \bibfield  {author} {\bibinfo {author} {\bibfnamefont {J.~P.~W.}\
  \bibnamefont {{Verbiest}}}, \bibinfo {author} {\bibfnamefont
  {L.}~\bibnamefont {{Lentati}}}, \bibinfo {author} {\bibfnamefont
  {G.}~\bibnamefont {{Hobbs}}}, \bibinfo {author} {\bibfnamefont
  {R.}~\bibnamefont {{van Haasteren}}}, \bibinfo {author} {\bibfnamefont
  {P.~B.}\ \bibnamefont {{Demorest}}}, \bibinfo {author} {\bibfnamefont
  {G.~H.}\ \bibnamefont {{Janssen}}}, \bibinfo {author} {\bibfnamefont {J.-B.}\
  \bibnamefont {{Wang}}}, \bibinfo {author} {\bibfnamefont {G.}~\bibnamefont
  {{Desvignes}}}, \bibinfo {author} {\bibfnamefont {R.~N.}\ \bibnamefont
  {{Caballero}}}, \bibinfo {author} {\bibfnamefont {M.~J.}\ \bibnamefont
  {{Keith}}}, \bibinfo {author} {\bibfnamefont {D.~J.}\ \bibnamefont
  {{Champion}}}, \emph {et~al.},\ }\href {https://doi.org/10.1093/mnras/stw347}
  {\bibfield  {journal} {\bibinfo  {journal} {\mnras}\ }\textbf {\bibinfo
  {volume} {458}},\ \bibinfo {pages} {1267} (\bibinfo {year} {2016})},\ \Eprint
  {https://arxiv.org/abs/1602.03640} {arXiv:1602.03640 [astro-ph.IM]}
  \BibitemShut {NoStop}%
\bibitem [{\citenamefont {{Graham}}\ \emph {et~al.}(2015)\citenamefont
  {{Graham}}, \citenamefont {{Djorgovski}}, \citenamefont {{Stern}},
  \citenamefont {{Drake}}, \citenamefont {{Mahabal}}, \citenamefont
  {{Donalek}}, \citenamefont {{Glikman}}, \citenamefont {{Larson}},\ and\
  \citenamefont {{Christensen}}}]{GrahamDjorgovski2015}%
  \BibitemOpen
  \bibfield  {author} {\bibinfo {author} {\bibfnamefont {M.~J.}\ \bibnamefont
  {{Graham}}}, \bibinfo {author} {\bibfnamefont {S.~G.}\ \bibnamefont
  {{Djorgovski}}}, \bibinfo {author} {\bibfnamefont {D.}~\bibnamefont
  {{Stern}}}, \bibinfo {author} {\bibfnamefont {A.~J.}\ \bibnamefont
  {{Drake}}}, \bibinfo {author} {\bibfnamefont {A.~A.}\ \bibnamefont
  {{Mahabal}}}, \bibinfo {author} {\bibfnamefont {C.}~\bibnamefont
  {{Donalek}}}, \bibinfo {author} {\bibfnamefont {E.}~\bibnamefont
  {{Glikman}}}, \bibinfo {author} {\bibfnamefont {S.}~\bibnamefont
  {{Larson}}},\ and\ \bibinfo {author} {\bibfnamefont {E.}~\bibnamefont
  {{Christensen}}},\ }\href {https://doi.org/10.1093/mnras/stv1726} {\bibfield
  {journal} {\bibinfo  {journal} {\mnras}\ }\textbf {\bibinfo {volume} {453}},\
  \bibinfo {pages} {1562} (\bibinfo {year} {2015})},\ \Eprint
  {https://arxiv.org/abs/1507.07603} {arXiv:1507.07603 [astro-ph.GA]}
  \BibitemShut {NoStop}%
\bibitem [{\citenamefont {{O'Neill}}\ \emph {et~al.}(2022)\citenamefont
  {{O'Neill}}, \citenamefont {{Kiehlmann}}, \citenamefont {{Readhead}},
  \citenamefont {{Aller}}, \citenamefont {{Blandford}}, \citenamefont
  {{Liodakis}}, \citenamefont {{Lister}}, \citenamefont {{Mr{\'o}z}},
  \citenamefont {{O'Dea}}, \citenamefont {{Pearson}}, \citenamefont {{Ravi}},
  \citenamefont {{Vallisneri}}, \citenamefont {{Cleary}}, \citenamefont
  {{Graham}}, \citenamefont {{Grainge}}, \citenamefont {{Hodges}},
  \citenamefont {{Hovatta}}, \citenamefont {{L{\"a}hteenm{\"a}ki}},
  \citenamefont {{Lamb}}, \citenamefont {{Lazio}}, \citenamefont
  {{Max-Moerbeck}}, \citenamefont {{Pavlidou}}, \citenamefont {{Prince}},
  \citenamefont {{Reeves}}, \citenamefont {{Tornikoski}}, \citenamefont
  {{Vergara de la Parra}},\ and\ \citenamefont
  {{Zensus}}}]{O'NeillKiehlmann2022}%
  \BibitemOpen
  \bibfield  {author} {\bibinfo {author} {\bibfnamefont {S.}~\bibnamefont
  {{O'Neill}}}, \bibinfo {author} {\bibfnamefont {S.}~\bibnamefont
  {{Kiehlmann}}}, \bibinfo {author} {\bibfnamefont {A.~C.~S.}\ \bibnamefont
  {{Readhead}}}, \bibinfo {author} {\bibfnamefont {M.~F.}\ \bibnamefont
  {{Aller}}}, \bibinfo {author} {\bibfnamefont {R.~D.}\ \bibnamefont
  {{Blandford}}}, \bibinfo {author} {\bibfnamefont {I.}~\bibnamefont
  {{Liodakis}}}, \bibinfo {author} {\bibfnamefont {M.~L.}\ \bibnamefont
  {{Lister}}}, \bibinfo {author} {\bibfnamefont {P.}~\bibnamefont
  {{Mr{\'o}z}}}, \bibinfo {author} {\bibfnamefont {C.~P.}\ \bibnamefont
  {{O'Dea}}}, \bibinfo {author} {\bibfnamefont {T.~J.}\ \bibnamefont
  {{Pearson}}}, \bibinfo {author} {\bibfnamefont {V.}~\bibnamefont {{Ravi}}},
  \bibinfo {author} {\bibfnamefont {M.}~\bibnamefont {{Vallisneri}}}, \bibinfo
  {author} {\bibfnamefont {K.~A.}\ \bibnamefont {{Cleary}}}, \bibinfo {author}
  {\bibfnamefont {M.~J.}\ \bibnamefont {{Graham}}}, \bibinfo {author}
  {\bibfnamefont {K.~J.~B.}\ \bibnamefont {{Grainge}}}, \bibinfo {author}
  {\bibfnamefont {M.~W.}\ \bibnamefont {{Hodges}}}, \bibinfo {author}
  {\bibfnamefont {T.}~\bibnamefont {{Hovatta}}}, \bibinfo {author}
  {\bibfnamefont {A.}~\bibnamefont {{L{\"a}hteenm{\"a}ki}}}, \bibinfo {author}
  {\bibfnamefont {J.~W.}\ \bibnamefont {{Lamb}}}, \bibinfo {author}
  {\bibfnamefont {T.~J.~W.}\ \bibnamefont {{Lazio}}}, \bibinfo {author}
  {\bibfnamefont {W.}~\bibnamefont {{Max-Moerbeck}}}, \bibinfo {author}
  {\bibfnamefont {V.}~\bibnamefont {{Pavlidou}}}, \bibinfo {author}
  {\bibfnamefont {T.~A.}\ \bibnamefont {{Prince}}}, \bibinfo {author}
  {\bibfnamefont {R.~A.}\ \bibnamefont {{Reeves}}}, \bibinfo {author}
  {\bibfnamefont {M.}~\bibnamefont {{Tornikoski}}}, \bibinfo {author}
  {\bibfnamefont {P.}~\bibnamefont {{Vergara de la Parra}}},\ and\ \bibinfo
  {author} {\bibfnamefont {J.~A.}\ \bibnamefont {{Zensus}}},\ }\href
  {https://doi.org/10.3847/2041-8213/ac504b} {\bibfield  {journal} {\bibinfo
  {journal} {\apjl}\ }\textbf {\bibinfo {volume} {926}},\ \bibinfo {eid} {L35}
  (\bibinfo {year} {2022})},\ \Eprint {https://arxiv.org/abs/2111.02436}
  {arXiv:2111.02436 [astro-ph.HE]} \BibitemShut {NoStop}%
\bibitem [{\citenamefont {{de la Parra}}\ \emph {et~al.}(2025)\citenamefont
  {{de la Parra}}, \citenamefont {{Kiehlmann}}, \citenamefont {{Mr{\'o}z}},
  \citenamefont {{Readhead}}, \citenamefont {{Synani}}, \citenamefont
  {{Begelman}}, \citenamefont {{Blandford}}, \citenamefont {{Ding}},
  \citenamefont {{Harrison}}, \citenamefont {{Liodakis}}, \citenamefont
  {{Max-Moerbeck}}, \citenamefont {{Pavlidou}}, \citenamefont {{Reeves}},
  \citenamefont {{Vallisneri}}, \citenamefont {{Aller}}, \citenamefont
  {{Graham}}, \citenamefont {{Hovatta}}, \citenamefont {{Lawrence}},
  \citenamefont {{Lazio}}, \citenamefont {{Mahabal}}, \citenamefont {{Molina}},
  \citenamefont {{O'Neill}}, \citenamefont {{Pearson}}, \citenamefont {{Ravi}},
  \citenamefont {{Tassis}},\ and\ \citenamefont
  {{Zensus}}}]{delaParraKiehlmann2025}%
  \BibitemOpen
  \bibfield  {author} {\bibinfo {author} {\bibfnamefont {P.~V.}\ \bibnamefont
  {{de la Parra}}}, \bibinfo {author} {\bibfnamefont {S.}~\bibnamefont
  {{Kiehlmann}}}, \bibinfo {author} {\bibfnamefont {P.}~\bibnamefont
  {{Mr{\'o}z}}}, \bibinfo {author} {\bibfnamefont {A.~C.~S.}\ \bibnamefont
  {{Readhead}}}, \bibinfo {author} {\bibfnamefont {A.}~\bibnamefont
  {{Synani}}}, \bibinfo {author} {\bibfnamefont {M.~C.}\ \bibnamefont
  {{Begelman}}}, \bibinfo {author} {\bibfnamefont {R.~D.}\ \bibnamefont
  {{Blandford}}}, \bibinfo {author} {\bibfnamefont {Y.}~\bibnamefont {{Ding}}},
  \bibinfo {author} {\bibfnamefont {F.}~\bibnamefont {{Harrison}}}, \bibinfo
  {author} {\bibfnamefont {I.}~\bibnamefont {{Liodakis}}}, \bibinfo {author}
  {\bibfnamefont {W.}~\bibnamefont {{Max-Moerbeck}}}, \bibinfo {author}
  {\bibfnamefont {V.}~\bibnamefont {{Pavlidou}}}, \bibinfo {author}
  {\bibfnamefont {R.}~\bibnamefont {{Reeves}}}, \bibinfo {author}
  {\bibfnamefont {M.}~\bibnamefont {{Vallisneri}}}, \bibinfo {author}
  {\bibfnamefont {M.~F.}\ \bibnamefont {{Aller}}}, \bibinfo {author}
  {\bibfnamefont {M.~J.}\ \bibnamefont {{Graham}}}, \bibinfo {author}
  {\bibfnamefont {T.}~\bibnamefont {{Hovatta}}}, \bibinfo {author}
  {\bibfnamefont {C.~R.}\ \bibnamefont {{Lawrence}}}, \bibinfo {author}
  {\bibfnamefont {T.~J.~W.}\ \bibnamefont {{Lazio}}}, \bibinfo {author}
  {\bibfnamefont {A.~A.}\ \bibnamefont {{Mahabal}}}, \bibinfo {author}
  {\bibfnamefont {B.}~\bibnamefont {{Molina}}}, \bibinfo {author}
  {\bibfnamefont {S.}~\bibnamefont {{O'Neill}}}, \bibinfo {author}
  {\bibfnamefont {T.~J.}\ \bibnamefont {{Pearson}}}, \bibinfo {author}
  {\bibfnamefont {V.}~\bibnamefont {{Ravi}}}, \bibinfo {author} {\bibfnamefont
  {K.}~\bibnamefont {{Tassis}}},\ and\ \bibinfo {author} {\bibfnamefont
  {J.~A.}\ \bibnamefont {{Zensus}}},\ }\href
  {https://doi.org/10.3847/1538-4357/addc60} {\bibfield  {journal} {\bibinfo
  {journal} {\apj}\ }\textbf {\bibinfo {volume} {987}},\ \bibinfo {eid} {191}
  (\bibinfo {year} {2025})},\ \Eprint {https://arxiv.org/abs/2408.02645}
  {arXiv:2408.02645 [astro-ph.HE]} \BibitemShut {NoStop}%
\bibitem [{\citenamefont {{Sudou}}\ \emph {et~al.}(2003)\citenamefont
  {{Sudou}}, \citenamefont {{Iguchi}}, \citenamefont {{Murata}},\ and\
  \citenamefont {{Taniguchi}}}]{SudouIguchi2003}%
  \BibitemOpen
  \bibfield  {author} {\bibinfo {author} {\bibfnamefont {H.}~\bibnamefont
  {{Sudou}}}, \bibinfo {author} {\bibfnamefont {S.}~\bibnamefont {{Iguchi}}},
  \bibinfo {author} {\bibfnamefont {Y.}~\bibnamefont {{Murata}}},\ and\
  \bibinfo {author} {\bibfnamefont {Y.}~\bibnamefont {{Taniguchi}}},\ }\href
  {https://doi.org/10.1126/science.1082817} {\bibfield  {journal} {\bibinfo
  {journal} {Science}\ }\textbf {\bibinfo {volume} {300}},\ \bibinfo {pages}
  {1263} (\bibinfo {year} {2003})},\ \Eprint
  {https://arxiv.org/abs/astro-ph/0306103} {arXiv:astro-ph/0306103 [astro-ph]}
  \BibitemShut {NoStop}%
\bibitem [{\citenamefont {{Hazboun}}\ \emph {et~al.}(2019)\citenamefont
  {{Hazboun}}, \citenamefont {{Romano}},\ and\ \citenamefont
  {{Smith}}}]{HazbounRomano2019}%
  \BibitemOpen
  \bibfield  {author} {\bibinfo {author} {\bibfnamefont {J.~S.}\ \bibnamefont
  {{Hazboun}}}, \bibinfo {author} {\bibfnamefont {J.~D.}\ \bibnamefont
  {{Romano}}},\ and\ \bibinfo {author} {\bibfnamefont {T.~L.}\ \bibnamefont
  {{Smith}}},\ }\href {https://doi.org/10.1103/PhysRevD.100.104028} {\bibfield
  {journal} {\bibinfo  {journal} {\prd}\ }\textbf {\bibinfo {volume} {100}},\
  \bibinfo {eid} {104028} (\bibinfo {year} {2019})},\ \Eprint
  {https://arxiv.org/abs/1907.04341} {arXiv:1907.04341 [gr-qc]} \BibitemShut
  {NoStop}%
\bibitem [{\citenamefont {{Nanograv Collaboration}}(2026)}]{NG15_targeted}%
  \BibitemOpen
  \bibfield  {author} {\bibinfo {author} {\bibnamefont {{Nanograv
  Collaboration}}},\ }\href {https://doi.org/10.3847/2041-8213/ae3719}
  {\bibfield  {journal} {\bibinfo  {journal} {\apjl}\ }\textbf {\bibinfo
  {volume} {998}},\ \bibinfo {eid} {L11} (\bibinfo {year} {2026})},\ \Eprint
  {https://arxiv.org/abs/2508.16534} {arXiv:2508.16534 [astro-ph.HE]}
  \BibitemShut {NoStop}%
\bibitem [{\citenamefont {{Gardiner}}\ \emph {et~al.}(2025)\citenamefont
  {{Gardiner}}, \citenamefont {{B{\'e}csy}}, \citenamefont {{Kelley}},\ and\
  \citenamefont {{Cornish}}}]{GardinerBecsy2025}%
  \BibitemOpen
  \bibfield  {author} {\bibinfo {author} {\bibfnamefont {E.~C.}\ \bibnamefont
  {{Gardiner}}}, \bibinfo {author} {\bibfnamefont {B.}~\bibnamefont
  {{B{\'e}csy}}}, \bibinfo {author} {\bibfnamefont {L.~Z.}\ \bibnamefont
  {{Kelley}}},\ and\ \bibinfo {author} {\bibfnamefont {N.~J.}\ \bibnamefont
  {{Cornish}}},\ }\href {https://doi.org/10.3847/1538-4357/ade4c2} {\bibfield
  {journal} {\bibinfo  {journal} {\apj}\ }\textbf {\bibinfo {volume} {988}},\
  \bibinfo {eid} {222} (\bibinfo {year} {2025})},\ \Eprint
  {https://arxiv.org/abs/2502.16016} {arXiv:2502.16016 [astro-ph.CO]}
  \BibitemShut {NoStop}%
\bibitem [{\citenamefont {{Goncharov}}\ \emph {et~al.}(2025)\citenamefont
  {{Goncharov}}, \citenamefont {{Sardana}}, \citenamefont {{Sesana}},
  \citenamefont {{Tomson}}, \citenamefont {{Antoniadis}}, \citenamefont
  {{Chalumeau}}, \citenamefont {{Champion}}, \citenamefont {{Chen}},
  \citenamefont {{Keane}}, \citenamefont {{Liu}}, \citenamefont {{Shaifullah}},
  \citenamefont {{Speri}},\ and\ \citenamefont
  {{Valtolina}}}]{GoncharovSardana2025a}%
  \BibitemOpen
  \bibfield  {author} {\bibinfo {author} {\bibfnamefont {B.}~\bibnamefont
  {{Goncharov}}}, \bibinfo {author} {\bibfnamefont {S.}~\bibnamefont
  {{Sardana}}}, \bibinfo {author} {\bibfnamefont {A.}~\bibnamefont {{Sesana}}},
  \bibinfo {author} {\bibfnamefont {S.~M.}\ \bibnamefont {{Tomson}}}, \bibinfo
  {author} {\bibfnamefont {J.}~\bibnamefont {{Antoniadis}}}, \bibinfo {author}
  {\bibfnamefont {A.}~\bibnamefont {{Chalumeau}}}, \bibinfo {author}
  {\bibfnamefont {D.}~\bibnamefont {{Champion}}}, \bibinfo {author}
  {\bibfnamefont {S.}~\bibnamefont {{Chen}}}, \bibinfo {author} {\bibfnamefont
  {E.~F.}\ \bibnamefont {{Keane}}}, \bibinfo {author} {\bibfnamefont
  {K.}~\bibnamefont {{Liu}}}, \bibinfo {author} {\bibfnamefont
  {G.}~\bibnamefont {{Shaifullah}}}, \bibinfo {author} {\bibfnamefont
  {L.}~\bibnamefont {{Speri}}},\ and\ \bibinfo {author} {\bibfnamefont
  {S.}~\bibnamefont {{Valtolina}}},\ }\href {https://doi.org/DOIL
  10.1038/s41467-025-65450-3} {\bibfield  {journal} {\bibinfo  {journal}
  {Nature Communications}\ }\textbf {\bibinfo {volume} {16}},\ \bibinfo {eid}
  {9692} (\bibinfo {year} {2025})},\ \Eprint {https://arxiv.org/abs/2409.03627}
  {arXiv:2409.03627 [astro-ph.HE]} \BibitemShut {NoStop}%
\bibitem [{\citenamefont {{McConnell}}\ and\ \citenamefont
  {{Ma}}(2013)}]{McConnellMa2013}%
  \BibitemOpen
  \bibfield  {author} {\bibinfo {author} {\bibfnamefont {N.~J.}\ \bibnamefont
  {{McConnell}}}\ and\ \bibinfo {author} {\bibfnamefont {C.-P.}\ \bibnamefont
  {{Ma}}},\ }\href {https://doi.org/10.1088/0004-637X/764/2/184} {\bibfield
  {journal} {\bibinfo  {journal} {\apj}\ }\textbf {\bibinfo {volume} {764}},\
  \bibinfo {eid} {184} (\bibinfo {year} {2013})},\ \Eprint
  {https://arxiv.org/abs/1211.2816} {arXiv:1211.2816 [astro-ph.CO]}
  \BibitemShut {NoStop}%
\bibitem [{\citenamefont {{Bernardi}}\ \emph {et~al.}(2010)\citenamefont
  {{Bernardi}}, \citenamefont {{Shankar}}, \citenamefont {{Hyde}},
  \citenamefont {{Mei}}, \citenamefont {{Marulli}},\ and\ \citenamefont
  {{Sheth}}}]{BernardiShankar2010}%
  \BibitemOpen
  \bibfield  {author} {\bibinfo {author} {\bibfnamefont {M.}~\bibnamefont
  {{Bernardi}}}, \bibinfo {author} {\bibfnamefont {F.}~\bibnamefont
  {{Shankar}}}, \bibinfo {author} {\bibfnamefont {J.~B.}\ \bibnamefont
  {{Hyde}}}, \bibinfo {author} {\bibfnamefont {S.}~\bibnamefont {{Mei}}},
  \bibinfo {author} {\bibfnamefont {F.}~\bibnamefont {{Marulli}}},\ and\
  \bibinfo {author} {\bibfnamefont {R.~K.}\ \bibnamefont {{Sheth}}},\ }\href
  {https://doi.org/10.1111/j.1365-2966.2010.16425.x} {\bibfield  {journal}
  {\bibinfo  {journal} {\mnras}\ }\textbf {\bibinfo {volume} {404}},\ \bibinfo
  {pages} {2087} (\bibinfo {year} {2010})},\ \Eprint
  {https://arxiv.org/abs/0910.1093} {arXiv:0910.1093 [astro-ph.CO]}
  \BibitemShut {NoStop}%
\bibitem [{\citenamefont {{Liepold}}\ and\ \citenamefont
  {{Ma}}(2024)}]{LiepoldMa2024}%
  \BibitemOpen
  \bibfield  {author} {\bibinfo {author} {\bibfnamefont {E.~R.}\ \bibnamefont
  {{Liepold}}}\ and\ \bibinfo {author} {\bibfnamefont {C.-P.}\ \bibnamefont
  {{Ma}}},\ }\href {https://doi.org/10.3847/2041-8213/ad66b8} {\bibfield
  {journal} {\bibinfo  {journal} {\apjl}\ }\textbf {\bibinfo {volume} {971}},\
  \bibinfo {eid} {L29} (\bibinfo {year} {2024})},\ \Eprint
  {https://arxiv.org/abs/2407.14595} {arXiv:2407.14595 [astro-ph.GA]}
  \BibitemShut {NoStop}%
\bibitem [{\citenamefont {{Sesana}}\ \emph {et~al.}(2014)\citenamefont
  {{Sesana}}, \citenamefont {{Barausse}}, \citenamefont {{Dotti}},\ and\
  \citenamefont {{Rossi}}}]{SesanaBarausse2014}%
  \BibitemOpen
  \bibfield  {author} {\bibinfo {author} {\bibfnamefont {A.}~\bibnamefont
  {{Sesana}}}, \bibinfo {author} {\bibfnamefont {E.}~\bibnamefont
  {{Barausse}}}, \bibinfo {author} {\bibfnamefont {M.}~\bibnamefont
  {{Dotti}}},\ and\ \bibinfo {author} {\bibfnamefont {E.~M.}\ \bibnamefont
  {{Rossi}}},\ }\href {https://doi.org/10.1088/0004-637X/794/2/104} {\bibfield
  {journal} {\bibinfo  {journal} {\apj}\ }\textbf {\bibinfo {volume} {794}},\
  \bibinfo {eid} {104} (\bibinfo {year} {2014})},\ \Eprint
  {https://arxiv.org/abs/1402.7088} {arXiv:1402.7088 [astro-ph.CO]}
  \BibitemShut {NoStop}%
\bibitem [{\citenamefont {{Falxa}}\ \emph {et~al.}(2025)\citenamefont
  {{Falxa}}, \citenamefont {{Leclere}},\ and\ \citenamefont
  {{Sesana}}}]{FalxaLeclere2025}%
  \BibitemOpen
  \bibfield  {author} {\bibinfo {author} {\bibfnamefont {M.}~\bibnamefont
  {{Falxa}}}, \bibinfo {author} {\bibfnamefont {H.~Q.}\ \bibnamefont
  {{Leclere}}},\ and\ \bibinfo {author} {\bibfnamefont {A.}~\bibnamefont
  {{Sesana}}},\ }\href {https://doi.org/10.1103/PhysRevD.111.023047} {\bibfield
   {journal} {\bibinfo  {journal} {\prd}\ }\textbf {\bibinfo {volume} {111}},\
  \bibinfo {eid} {023047} (\bibinfo {year} {2025})},\ \Eprint
  {https://arxiv.org/abs/2412.01899} {arXiv:2412.01899 [gr-qc]} \BibitemShut
  {NoStop}%
\bibitem [{\citenamefont {{Ravi}}\ \emph {et~al.}(2014)\citenamefont {{Ravi}},
  \citenamefont {{Wyithe}}, \citenamefont {{Shannon}}, \citenamefont
  {{Hobbs}},\ and\ \citenamefont {{Manchester}}}]{RaviWyithe2014}%
  \BibitemOpen
  \bibfield  {author} {\bibinfo {author} {\bibfnamefont {V.}~\bibnamefont
  {{Ravi}}}, \bibinfo {author} {\bibfnamefont {J.~S.~B.}\ \bibnamefont
  {{Wyithe}}}, \bibinfo {author} {\bibfnamefont {R.~M.}\ \bibnamefont
  {{Shannon}}}, \bibinfo {author} {\bibfnamefont {G.}~\bibnamefont {{Hobbs}}},\
  and\ \bibinfo {author} {\bibfnamefont {R.~N.}\ \bibnamefont {{Manchester}}},\
  }\href {https://doi.org/10.1093/mnras/stu779} {\bibfield  {journal} {\bibinfo
   {journal} {\mnras}\ }\textbf {\bibinfo {volume} {442}},\ \bibinfo {pages}
  {56} (\bibinfo {year} {2014})},\ \Eprint {https://arxiv.org/abs/1404.5183}
  {arXiv:1404.5183 [astro-ph.CO]} \BibitemShut {NoStop}%
\bibitem [{\citenamefont {{Sato-Polito}}\ \emph {et~al.}(2024)\citenamefont
  {{Sato-Polito}}, \citenamefont {{Zaldarriaga}},\ and\ \citenamefont
  {{Quataert}}}]{Sato-PolitoZaldarriaga2024}%
  \BibitemOpen
  \bibfield  {author} {\bibinfo {author} {\bibfnamefont {G.}~\bibnamefont
  {{Sato-Polito}}}, \bibinfo {author} {\bibfnamefont {M.}~\bibnamefont
  {{Zaldarriaga}}},\ and\ \bibinfo {author} {\bibfnamefont {E.}~\bibnamefont
  {{Quataert}}},\ }\href {https://doi.org/10.1103/PhysRevD.110.063020}
  {\bibfield  {journal} {\bibinfo  {journal} {\prd}\ }\textbf {\bibinfo
  {volume} {110}},\ \bibinfo {eid} {063020} (\bibinfo {year}
  {2024})}\BibitemShut {NoStop}%
\bibitem [{\citenamefont {{EPTA Collaboration}}\ and\ \citenamefont {{InPTA
  Collaboration}}(2024)}]{EPTA_DR2_NEWPHYS}%
  \BibitemOpen
  \bibfield  {author} {\bibinfo {author} {\bibnamefont {{EPTA Collaboration}}}\
  and\ \bibinfo {author} {\bibnamefont {{InPTA Collaboration}}},\ }\href
  {https://doi.org/10.1051/0004-6361/202347433} {\bibfield  {journal} {\bibinfo
   {journal} {\aap}\ }\textbf {\bibinfo {volume} {685}},\ \bibinfo {eid} {A94}
  (\bibinfo {year} {2024})},\ \Eprint {https://arxiv.org/abs/2306.16227}
  {arXiv:2306.16227 [astro-ph.CO]} \BibitemShut {NoStop}%
\bibitem [{\citenamefont {{Casey-Clyde}}\ \emph {et~al.}(2025)\citenamefont
  {{Casey-Clyde}}, \citenamefont {{Mingarelli}}, \citenamefont {{Greene}},
  \citenamefont {{Goulding}}, \citenamefont {{Chen}},\ and\ \citenamefont
  {{Trump}}}]{Casey-ClydeMingarelli2025}%
  \BibitemOpen
  \bibfield  {author} {\bibinfo {author} {\bibfnamefont {J.~A.}\ \bibnamefont
  {{Casey-Clyde}}}, \bibinfo {author} {\bibfnamefont {C.~M.~F.}\ \bibnamefont
  {{Mingarelli}}}, \bibinfo {author} {\bibfnamefont {J.~E.}\ \bibnamefont
  {{Greene}}}, \bibinfo {author} {\bibfnamefont {A.~D.}\ \bibnamefont
  {{Goulding}}}, \bibinfo {author} {\bibfnamefont {S.}~\bibnamefont {{Chen}}},\
  and\ \bibinfo {author} {\bibfnamefont {J.~R.}\ \bibnamefont {{Trump}}},\
  }\href {https://doi.org/10.3847/1538-4357/adce05} {\bibfield  {journal}
  {\bibinfo  {journal} {\apj}\ }\textbf {\bibinfo {volume} {987}},\ \bibinfo
  {eid} {106} (\bibinfo {year} {2025})},\ \Eprint
  {https://arxiv.org/abs/2405.19406} {arXiv:2405.19406 [astro-ph.HE]}
  \BibitemShut {NoStop}%
\bibitem [{\citenamefont {{Sesana}}\ \emph {et~al.}(2018)\citenamefont
  {{Sesana}}, \citenamefont {{Haiman}}, \citenamefont {{Kocsis}},\ and\
  \citenamefont {{Kelley}}}]{SesanaHaiman2018}%
  \BibitemOpen
  \bibfield  {author} {\bibinfo {author} {\bibfnamefont {A.}~\bibnamefont
  {{Sesana}}}, \bibinfo {author} {\bibfnamefont {Z.}~\bibnamefont {{Haiman}}},
  \bibinfo {author} {\bibfnamefont {B.}~\bibnamefont {{Kocsis}}},\ and\
  \bibinfo {author} {\bibfnamefont {L.~Z.}\ \bibnamefont {{Kelley}}},\ }\href
  {https://doi.org/10.3847/1538-4357/aaad0f} {\bibfield  {journal} {\bibinfo
  {journal} {\apj}\ }\textbf {\bibinfo {volume} {856}},\ \bibinfo {eid} {42}
  (\bibinfo {year} {2018})},\ \Eprint {https://arxiv.org/abs/1703.10611}
  {arXiv:1703.10611 [astro-ph.HE]} \BibitemShut {NoStop}%
\end{thebibliography}%
\bibliographystyle{apsrev4-2}

\appendix

\section{\label{app:strain_units}Units of GW strain}

In Section~\ref{sec:model} and the rest of the manuscript, we operate primarily with two physical units related to GW strain. 
First, the strain amplitude $h_0$ of a monochromatic CW signal produced by an SMBHB. 
Second, the characteristic strain, which stems from the power spectral density and is thus specific to stochastic sources. 
In this Section, we expand on the relation of these units through Equation~\ref{eq:cw_h0_hcw}, providing a broader overview of the strain-related quantities in GW astronomy. 

The dimensionless \textit{strain amplitude} $h_0$ of a monochromatic CW signal with a frequency $f_{\rm 0}$, given by the first equality in Equation~\ref{eq:cw_h0_hcw}. 
In some literature, the factor of $2$ is omitted. 
It is such that the time-domain GW strain is
\begin{equation}
  \begin{cases}
    h_+(t) = h_0 (1 + \cos^2 \iota) \cos \Phi(t),\\
    h_\times(t) = -2 h_0 \cos \iota \sin \Phi(t), 
  \end{cases}
\end{equation}
where $\iota$ is the inclination angle of the binary with respect to the line of sight, $\Phi(t)$ is the GW phase time series, and $(+,\times)$ are the tensor GW polarizations of General Relativity. 

To obtain the second equality in Equation~\ref{eq:cw_h0_hcw}, we must first introduce the concept of the root-mean-squared \textit{(RMS) strain amplitude}. 
It is the strain amplitude averaged over binary orientations, 
\begin{equation}
  h^2_{\rm rms} = \langle h^2_+ + h^2_\times \rangle_{\Phi,\iota}
\end{equation}
A uniform distribution on the $(\iota,\Phi)$ sphere corresponds to the uniform distribution in $\cos \iota$ and the uniform distribution in $\Phi$. 
Based on the above, $\langle \cos^2 \iota \rangle|_{\mathcal{U}(\cos \iota)} = 1/3$, and $\langle \cos^4 \iota \rangle|_{\mathcal{U}(\cos \iota)} = 1/5$. 
Whereas $\langle \cos^2 \Phi \rangle|_{\mathcal{U}(\Phi)} = \langle \sin^2 \Phi \rangle|_{\mathcal{U}(\Phi)} = 1/2$.
Proceeding with the averaging yields 
\begin{equation}
  h_{\rm rms} = \frac{1}{2} \sqrt{\frac{32}{5}} h_0.
  \label{eq:h_rms_to_h0}
\end{equation}

The next useful quantity is the \textit{Fourier-domain strain}, $\tilde{h}(f)$ in units [s], 
\begin{equation}
  \tilde{h}(f) = \int h(t) e^{-2\pi f t} dt,
\end{equation}
where the limits of the integration represent the observation time. 
For the infinite observation time of the monochromatic signal at $f_0$, $\tilde{h}(f_0)=\delta{f_0}$. 
The \textit{RMS Fourier-domain strain} is then 
\begin{equation}
  \sqrt{\langle |\tilde{h}(f_0)|^2 \rangle} = \sqrt{\langle |\tilde{h}_+(f_0)|^2 + |\tilde{h}_\times(f_0)|^2 \rangle},
\end{equation}
where averaging is performed again over $\mathcal{U}(\cos \iota)$. 
For the finite observation span $T_{\rm obs}$, computing the integral analytically and averaging over $\cos \iota$ as in the previous paragraph, yields
\begin{equation}
  \sqrt{\langle |\tilde{h}(f_0)|^2 \rangle}_{\mathcal{U}(\cos \iota)} = \frac{T_{\rm obs}}{\sqrt{2}} h_{\rm rms} \bigg |_{T_{\rm obs} \gg f_0^{-1}}.
\end{equation}

The characteristic strain is defined as $\sqrt{S(f) f}$, whereas the one-sided strain power spectral density is defined as
\begin{equation}
  S(f) \equiv \lim_{T_{\rm obs} \rightarrow \infty} 2 \frac{|\tilde{h}(f)|^2}{T_{\rm obs}}. 
\end{equation}
Therefore, the characteristic strain is calculated from the Fourier-domain RMS strain in the limit of the large observation span, 
\begin{equation}
  h^2(f) \approx 2 \frac{\langle |\tilde{h}(f_0)|^2 \rangle}{T_{\rm obs}} f = \frac{2 h^2_{\rm rms} f}{T_{\rm obs}} \frac{T^2_{\rm obs}}{2}  = h^2_{\rm rms} f T_{\rm obs}.  
  \label{eq:h_rms_to_c}
\end{equation}
It can be further expanded as the right-side equality in Equation~\ref{eq:cw_h0_hcw}, $h^2_{\rm rms}f/\Delta f$, where $\Delta f = T_{\rm obs}^{-1}$. 
The same calculation applies to other quantities in dimensionless units of characteristic strain: $h_{\rm s}$, $h_{\rm t}$, $h_{\rm t-1}$, $h_{\rm cw}$. 

\section{\label{app:model}The Astrophysical Model of Nanohertz Gravitational Waves}

In this work, we model the distribution of discrete GW sources similarly to Ref.~\cite{Sato-PolitoZaldarriaga2025}. Below, we summarize some of the main results used to model the distribution of the GWB. We then extend the formalism introduced in \cite{Sato-PolitoZaldarriaga2025} to model the distribution of the brightest source, which corresponds to the continuous-wave amplitude.

\subsection{\label{app:model:strain_count} Strain number-count}

The characteristic strain at a frequency bin of width $\Delta f$ measured by a PTA is given by the sum of the contributions of each binary SMBH emitting in that band. The inclination and polarization-averaged strain that a circular binary with chirp mass $\mathcal{M}$ radiating at a rest-frequency $f_r$ at a redshift $z$ contributes is given by
\begin{equation}
\begin{split}
    \tilde h^2 (\mathcal{M},z,f)= \\ = \frac{32}{5} \left(\frac{G \mathcal{M}}{c^2}\right)^{10/3} \frac{(1+z)^{4/3}}{\chi^2(z)} \left(\frac{\pi f}{c}\right)^{4/3} \frac{f}{\Delta f},
\end{split}
    \label{eq:tildehs2}
\end{equation}
where $\chi$ is the comoving radial distance and $f=f_r/(1+z)$ is the observed frequency. 
The total number of SMBHBs, characterized by a set of parameters $\vec{\theta} = (\mathcal{M},z)$ can be related to the comoving merger rate density $dn/d\vec{\theta}$ by
\begin{equation}
    \frac{dN}{d\vec{\theta} d\log f} = \frac{dn}{d\vec{\theta}} \frac{dt_r}{d\log f} \frac{dz}{dt_r} \frac{dV_c}{dz}.
    \label{eq:dNdlogh2}
\end{equation}
where $t_r$ is the coordinate time in the source rest frame and $V_c$ is the comoving volume. We will assume that the mass and redshift dependence of the population distribution is separable, so that $\frac{dn}{d\vec{\theta}} \equiv \frac{dn}{d\mathcal{M}} p(z)$. The frequency evolution is given by
\begin{equation}
    \frac{d\log f_r}{dt_r} = \frac{96 \pi^{8/3}}{5}\left(\frac{G \mathcal{M}_c}{c^3}\right)^{5/3} f_r^{8/3},
    \label{eq:dlogfdt}
\end{equation}
and
\begin{equation}
    \frac{dz}{dt_r} \frac{dV_c}{dz} = (1+z)H(z) \frac{4\pi c \chi^2(z)}{H(z)}.
    \label{eq:dVdt}
\end{equation}
Hence, the volume factor on the right-hand side is 
\begin{equation}
\frac{dV_c}{d\log f} \propto \mathcal{M}^{-5/3} f^{-8/3} \chi(z)^2(1+z)^{-5/3}.
\end{equation}

The crucial quantity for predicting the probability distribution of both the power in the GWB and the brightest source is the number of sources per strain amplitude weighted by the strain $h^2_s dN/\log h^2_s d\log f$, which can be computed from Eq.~\ref{eq:dNdlogh2} as
\begin{equation}
\begin{split}
    h^2_s \frac{dN}{d\log h^2_s d\log f} = \\ = \int \tilde h^2(\vec{\theta}) \frac{dN}{d\vec{\theta} d\log f}\delta[\log h^2_s - \log \tilde h^2(\vec{\theta})] d\vec{\theta}\ .
\end{split}
\label{eq:h2s_dNdlogh2s}
\end{equation}

\subsection{\label{app:model:mu_x_interp} Approach 1 towards the calculation of $\mu(x)$: interpolation}
In this work, we calculate $\mu(x)$ based on selecting an initial set of astrophysical parameters. 
The luminosity function is initially calculated based on Equation~\ref{eq:galaxy_property_scaling}, Equation~\ref{eq:sigma_function}, and the strain number count defined in Equation~\ref{eq:h2s_dNdlogh2s}. 
This calculation necessitates PDFs of the mass ratio $q$ and redshift $z$ of SMBHBs. 
We define them as 
\begin{equation}
    p(q) = q^{\delta}
    \label{eq:mass_ratio_pdf}
\end{equation}
and
\begin{equation}
    p(z) = z^\gamma e^{-z/z_*},
    \label{eq:redshift_pdf}
\end{equation}
respectively. 
To calculate a specific luminosity function and thus $\mu(x)$, we need to fix the parameters of the aforementioned equations. 
We chose $\epsilon_0=0.38$, $\phi_*=2.611 \times 10^{-2}~[{\rm Mpc}^{-3}]$, $\sigma_*=170~[{\rm km/s}]$, $\alpha=0.41$, $\beta=2.59$, $a_\bullet=8.32~[\log_{10}M_\odot]$, $b_\bullet=5.64~[\log_{10}M_\odot]$, $\delta=-1$, $\gamma=0.5$, $z_*=0.33$ based on Ref.~\cite{Sato-PolitoZaldarriaga2025}. 
Effectively, these are astrophysical hyperparameters which are reduced to $(N_{\rm c},h_{\rm c})$ due to the invariance of $\mu(x)$. 
A linear interpolation allows us to calculate $\mu(x)$ for any $x$, not only the initial set of points. 

\subsection{\label{app:model:mu_x_logNormal} Approach 2 towards the calculation of $\mu(x)$: the log-Normal approximation}

Although the integral of Equation~\ref{eq:h2s_dNdlogh2s} can be evaluated numerically for any SMBHB population model of choice, it is useful for future work to derive a more general parametrization by using a saddle-point approximation. 
Expressing the integrand as
\begin{equation}
    \tilde{h}^2(\vec{\theta}) \frac{dN}{d\vec{\theta} d\log f} \equiv e^{-g(\vec{\theta}, f)}
\end{equation}
and expanding $g(\vec{\theta}, f)$ around the peak $\vec{\theta}_\star = (\log \mathcal{M}_{\star}, z_\star)$, leads to
\begin{equation}
    g(m, z, f) \approx g_\star(f) + \frac{1}{2} \frac{(m - m_\star)^2}{\sigma^2_m} + \frac{1}{2} \frac{(z - z_\star)^2}{\sigma^2_z},
\end{equation}
where we defined $\log \mathcal{M} = m$, $g_\star(f) = g(m_\star, z_\star, f)$, and the variances $\frac{\partial^2 g}{\partial \theta_i^2}(m_\star, z_\star, f) = \sigma^{-2}_i$ for $i = m, z$ for brevity. Substituting into Eq.~\ref{eq:h2s_dNdlogh2s}, we find that
\begin{equation}
    \begin{split}
    h^2_s \frac{dN}{d\log h^2_s d\log f} \approx e^{-g_\star(f)} \times \\ 
    \times \int  \exp\Bigg\{-\frac{1}{2} \frac{(m - m_\star)^2}{\sigma^2_m}  - \frac{1}{2} \frac{(z - z_\star)^2}{\sigma^2_z}\Bigg\} \times \\ 
    \times \delta[\log h^2_s - \log \tilde h^2(m,z,f)] dm dz .
    \end{split}
    \label{eq:h2s_dNdlogh2s_saddle}
\end{equation}
Expanding $\log \tilde h^2(m,z,f)$ to first order around the peak, we have that
\begin{equation}
\begin{split}
    \log \tilde h^2\equiv \ \frac{10}{3}m + \Phi(z) + \frac{7}{3} \log f + {\rm const} \approx \\ \approx \log \tilde h^2_{s, \rm peak}(f) + \frac{10}{3} (m - m_\star) + \\ + \Phi'(z_\star) (z - z_\star),
\end{split}
\end{equation}
where the first equality defines the redshift-dependent term $\Phi(z) \equiv \tfrac{4}{3}\log (1+z) - 2 \log \chi $. We can now compute the integrals in Eq.~\ref{eq:h2s_dNdlogh2s_saddle} directly
\begin{equation}
\begin{split}
    h^2_s \frac{dN}{d\log h^2_s d\log f} \approx \frac{h^2_c(f)}{\sqrt{2\pi} \sigma} \times \\ \times \exp\left\{-\frac{1}{2} \frac{(\log h^2_s - \log h^2_{s, \rm peak})^2}{\sigma^2}\right\},
\end{split}
\end{equation}
where $h^2_c(f) = \int d\log h^2_s \tfrac{dN}{d\log h^2_s d\log f} h^2_s$ is the mean characteristic strain amplitude of the population, and we defined
\begin{equation}
    \sigma^2 = \left(\frac{10}{3}\right)^2 \sigma^2_m + \left(\Phi'(z_\star)\sigma_z\right)^2.
\end{equation}
In summary, we approximate the distributions of $\log \mathcal{M}$ and $z$ by Gaussians around the location of the maximum of the characteristic strain contribution. By expanding $\log h^2$ to first order around the peak, we can then express $\log h^2$ as a sum of Gaussian variables, and therefore $h^2_s dN/d\log h^2_s$ is also a Gaussian.

We define a rescaled strain variable as $x=h^2_s/h^2_{s, \rm peak}$. The number-count of sources per log rescaled characteristic strain is then given by
\begin{equation}
    \begin{split}
    \frac{dN}{d\log x d\log f} \approx \frac{h^2_c}{h^2_{s, \rm peak}} \frac{1}{\sqrt{2\pi} \sigma x} \times \\ \times \exp\left\{-\frac{1}{2} \frac{(\log x)^2}{\sigma^2}\right\} \equiv \\ 
    \equiv N_c(f) \mu(x),
    \end{split}
    \label{eq:dNdxdf_saddle}
\end{equation}
where the second line defines the log-normal shape $\mu(x)$ and the characteristic number of sources $N_c(f) = h^2_c/h^2_{s, \rm peak}$. 

With the approximations employed in this section, we have derived a general expression for astrophysical strain distributions in terms of three effective parameters: the strain that most contributes to GWB amplitude $h^2_{s, {\rm peak}}$, the characteristic number of source $N_c$ which corresponds to the number of sources of strain $h^2_{s, {\rm peak}}$ required to produce a total amplitude $h^2_c$, and the variance $\sigma^2$. 
The variance $\sigma^2$ is to be fixed to an astrophysically motivated value, based on galaxy velocity dispersion functions and scaling relations as per Section~\ref{app:model:mu_x_interp}. 
The strain distribution across all frequencies is therefore fully specified by 2 effective parameters. 
We refer the reader to Refs.~\cite{Sato-PolitoZaldarriaga2024} and \cite{Sato-PolitoZaldarriaga2025} for a more in-depth discussion of the relation between the effective parameters and astrophysical parametrizations of the SMBH merger rate density.

\subsection{GWB distribution}
The observed GWB for a discrete set of sources is given by the sum of characteristic strain amplitudes of all sources emitting in a particular frequency bin. That is, in terms of the rescaled characteristic strain
\begin{equation}
    x_t = \sum_s x_s,
\end{equation}
where $s$ labels each source.
The brightness of each source follows the distribution given in Eq.~\ref{eq:dNdxdf_saddle}, while the number of discrete sources is assumed to follow a Poisson distribution. As in Ref.~\cite{Sato-PolitoZaldarriaga2025}, the distribution of the total GWB amplitude is given by
\begin{equation}
    P(x_t) = \sum_{N_s = 0}^{\infty} P(x_t | N_s) P(N_s|\bar{N}),
    \label{eq:Pxt}
\end{equation}
where $P(N_s|\bar{N})$ is a Poisson distribution and $P(x_t | N_s)$ is the probability that $N_s$ sources produce a GWB amplitude that adds up to $x_t$. The latter is given by $N_s$ convolutions of the single-source distribution $P_1 (x) \equiv P(x|N_s=1) = \tfrac{1}{\bar{N}} \tfrac{dN}{dx}$ with itself, which can be more conveniently evaluated in the Fourier domain.
The characteristic function of the GWB amplitude is given by
\begin{equation}
    \tilde{P}(\omega) = \int dx_t \ P(x_t) e^{i \omega x_t},
    \label{eq:P_omega}
\end{equation}
and likewise for the single source PDF. As shown in \cite{Sato-PolitoZaldarriaga2025}, the Fourier transform of Eq.~\ref{eq:Pxt} can be expressed as
\begin{equation}
    \tilde{P}(\omega) = e^{u(\omega)},
\end{equation}
where
\begin{equation}
    \begin{split}
    u(\omega) = \bar{N} \left[\tilde{P}_1(\omega) - 1\right] = \\ = \int_0^{\infty} dx \ \frac{dN}{dx} \left[e^{i \omega x} - 1\right],
    \end{split}
\end{equation}
and the probability distribution of the GWB is recovered by taking the inverse Fourier transform.

\subsection{\label{app:p_hmax}PDF of the strain of the brightest source}
In a continuous-wave search, the imprint of the loudest binary on the timing residuals is modeled as a resolved point source. In this section, we compute the probability distribution $P_{\rm max}$ of the maximum strain, $x_{\rm cw} = \max_s x_s$, from the same astrophysical distribution as the GWB. 

Given a set of $N_s$ sources, the maximum is below $x$ if the brightness of each source is below $x$. Hence, the cumulative distribution function (CDF) $F(x) = \int_{-\infty}^{x} dx' P(x')$ of the maximum of $N_s$ independent and identically distributed sources is 
\begin{equation}
    F_{\rm max}(x|N_s) = \left[ F_1(x) \right]^{N_s},
\end{equation}
where $F_{1}$ and $F_{\rm max}$ are the CDFs of the single source and maximum brightness, respectively. Since the number of sources is Poisson-distributed, the CDF of the maximum is therefore
\begin{equation}
\begin{split}
    F_{\rm max}(x) = \sum_{N_s=0}^{\infty} \left[ F_1(x) \right]^{N_s} P(N_s|\bar{N}) = \\
    =\ e^{-\bar N} \sum_{N_s} \frac{\bar {N}^{N_s}}{N_s !} \left[ F_1(x) \right]^{N_s} = \\ = e^{-\bar{N} (1-F_{1}(x))},
\end{split}
\end{equation}
where in the first equality we substitute the expression for $F_{\rm max}(x|N_s)$ and in the third equality we identify the Poisson distribution and CDF with the exponential series. Since the exponent is $\bar{N} (1-F_{1}(x)) = \int_x^{\infty} dx' \tfrac{dN}{dx'}$, differentiating with respect to $x$ yields the PDF
\begin{equation}
    P_{\rm max} (x) = \frac{dN}{dx} \exp\left\{-\int^\infty_x dx' \frac{dN}{dx'} \right\}.
\end{equation}


\end{document}